\documentclass[aip,jcp,showpacs,showkeys,preprint,longbibliography]{revtex4-1}

\usepackage{amssymb}
\usepackage{amsmath}
\usepackage{graphicx}
\usepackage{amsbsy}
\usepackage{float}
\usepackage{color}

\newcommand{\bq}{\begin{eqnarray}}
\newcommand{\eq}{\end{eqnarray}}
\newcommand{\bqn}{\begin{eqnarray*}}
\newcommand{\eqn}{\end{eqnarray*}}
\newcommand{\nn}{{\bf n}}
\newcommand{\xx}{{\bf x}}
\newcommand{\rr}{{\bf r}}
\newcommand{\RR}{{\bf R}}
\newcommand{\kk}{{\bf k}}

\newcommand{\vv}{{\bf v}}
\newcommand{\LL}{{\bf L}}
\newcommand{\Mu}{\mu_c}
\newcommand{\SSS}{{\bf S}}
\newcommand{\sgn}{\text{sgn}}
\newcommand{\nablab}{\pmb{\nabla}}
\newcommand{\anew}[1]{{#1}}

\begin{document}
\title{Radial distribution function in a diffusion Monte Carlo
  simulation of a Fermion fluid between the ideal gas
  and the Jellium model}

\author{Riccardo Fantoni}
\email{rfantoni@ts.infn.it}
\affiliation{Dipartimento di Scienze dei Materiali e Nanosistemi,
  Universit\`a Ca' Foscari Venezia, Calle Larga S. Marta DD2137,
  I-30123 Venezia, Italy} 

\date{\today}

\begin{abstract}
We study, through the diffusion Monte Carlo method, a spin one-half
fermion fluid, in the three dimensional Euclidean space, at zero
temperature. The point particles, immersed in a uniform ``neutralizing''
background, interact with a pair-potential which can be continuously
changed from zero to the Coulomb potential depending on a parameter $\mu$. 
We determine the radial distribution functions
of the system for various values of density, $\mu$, and polarization.
We discuss about the importance, in a computer experiment, of the choice
of suitable estimators to measure a physical quantity. The radial
distribution function is determined through the usual histrogram
estimator and through an estimator determined via the use of the
Hellmann and Feynman theorem. In a diffusion Monte Carlo simulation
the latter route introduces a new bias to the measure of the radial
distribution function due to the choice of the auxiliary function.
This bias is independent from the usual one due to the choice of the
trial wave function. A brief account of the results from this study
were presented in a recent communication [R. Fantoni, Solid state
Communications, {\bf 159}, 106 (2013)].  
\end{abstract}


\maketitle
\section{Introduction}
\label{sec:introduction}
The Jellium model is a system of pointwise electrons of charge $e$ and
number density 
$n$ in the three dimensional Euclidean space filled with an uniform
neutralizing background of charge density $-en$. The zero
temperature, ground-sate, properties of the statistical mechanical
system thus depends just on the electronic density $n$ or the
Wigner-Seitz radius $r_s=(3/4\pi n)^{1/3}/a_0$ where
$a_0$ is Bohr radius. The model can be used for example as a first
approximation to describe free electrons in metallic elements
\anew{\cite{Ashcroft-Mermin}} ($2\lesssim r_s \lesssim 4$) or a white dwarf
\anew{\cite{Shapiro-Teukolsky}} ($r_s\simeq 0.01$). 

When an impurity of charge $q$ is added to the system, the screening
cloud of electrons will experience the Friedel oscillations. In the
Thomas-Fermi description of the static screening an electric potential
$qv_H(\rr)$ (the {\sl Hartree} potential) is created by the impurity
and by the redistribution of the electronic charge $n(\rr)-n$. It obeys
the Poisson equation $qe\nabla^2v_H(\rr)=4\pi
e[-q\delta(\rr)-en(\rr)+en]$ and the equilibrium condition on the
electrochemical potential, $\Mu(n(\rr))+qev_H(\rr)=\text{constant}$. An
analytic solution can be obtained for $|q|\ll 1$, when we find
$n(\rr)-n\simeq -qev_H(\rr)\partial n/\partial\Mu$ by expansion of $\Mu$
around the homogeneous state. Assuming $\partial n/\partial\Mu$ is
positive and with the definition $k_s=\sqrt{4\pi e^2\partial
  n/\partial\Mu}$ the Poisson equation yields
\bq
v_H(r)=\frac{e^{-k_sr}}{r}~.
\eq 
It is clear from this result that the quantity $1/k_s$ measures the
distance over which the self consistent potential associated with the
impurity penetrates into the electron gas. Thus, $1/k_s$ has the
meaning of a screening length. The Thomas-Fermi value of the screening
length is obtained by replacing the thermodynamic quantity $\partial
n/\partial\Mu$ by its value for non-interacting fermions, using for
$\Mu$ the Fermi energy. Clearly we have that $v_H(r)\to 1/r$ as
$1/k_s\to\infty$ and $v_H(r)\to 0$ as $1/k_s\to 0$. Also $v_H$ is
short ranged.

It is important to study the ground-state properties of a model of
point fermions of spin one-half interacting with a bare pair-potential
$v_\mu(r)$ which can be continuously changed from zero ($\mu\to 0$,
ideal gas) to the Coulomb potential ($\mu\to\infty$, Jellium model)
depending on a parameter $\mu$. And we chose the following functional
form   
\bq \label{Paz}
v_\mu(r)=\frac{\text{erf}(\mu r)}{r}~.
\eq 
Still the fluid is immersed in a static uniform background of continuously
distributed point particles which interact with the particles of the
fluid with the same pair-potential but of opposite sign.

A major challenge in the Kohn-Sham scheme of Density Functional Theory
is to devise approximations to the exchange-correlation functional
that accurately describes near-degeneracy or long-range correlation
effects such as van der Waals forces. Among recent progresses to
circumvent this problem, we mention ``range-separated'' density
functional schemes which combine the Kohn-Sham formalism with either
random-phase approximation \cite{Zhu2010} or multideterminantal
approaches \cite{Toulouse2005}. 
Such schemes require a local density functional for particles
interacting via modified potentials defined in terms of a suitable
parameter $\mu$ which either softens the core or suppresses the
long-range tail.  Further insight into electronic correlations in
molecules and materials can be gained through the analysis of the
on-top pair correlation function \cite{Gori2006}. 

\anew{Within Quantum Monte Carlo, the Diffusion Monte Carlo} is the
method of choice for the calculation of ground-state properties of
appropriate reference homogeneous systems, \anew{(the path integral method
\cite{Ceperley1995} can be used to extend the study to non-zero
temperatures degenerate systems \cite{Ceperley1996}),}  
the most relevant example being the correlation energy of the electron
gas obtained by Ceperley and Alder back in 1980
\cite{Ceperley1980}. This is even more 
so in the present days, since better wave-functions and optimization
methods have been developed, better schemes to minimize finite-size
effect have been devised, and vastly improved computational facilities
are available. 

Recently, Zecca {\sl et al.} \cite{Zecca2004}  have provided a Local
Density functional for short-range pair potentials $v(r)=\text{erfc}(\mu
r)/r$, whereas Paziani {\sl et al.} \cite{Paziani2006} have developed
a Local Spin Density functional for the softened-core, long range
case, $v(r)=\text{erf}(\mu r)/r$. 

It is the purpose of this work to build on previous work
\cite{Zecca2004,Paziani2006} and provide the Radial Distribution
Function (RDF), most notably the on-top value, {\sl i.e.} its value at
contact, at a zero radial distance, for the pair potential
of Ref. \onlinecite{Paziani2006}, given in Eq. (\ref{Paz}). A brief
account of the results from this study has been presented in a recent
communication \cite{Fantoni13}. Aim of the present work is to give a
complete and detailed account of the calculations that has been
carried on for such a study. 

We performed fixed-nodes Diffusion Monte Carlo simulations 
\cite{Kolorenc2011} where we used modern techniques \cite{Foulkes2001}
to optimize Slater-Jastrow wave-functions with backflow and three-body
correlations \cite{Kwon1998} and Hellmann and Feynman (HFM) measures
\cite{Toulouse2007} to calculate the RDF, particularly the on-top
value, which suffers from poor statistical sampling in its
conventional histogram implementation. Twist-averaged boundary
conditions \cite{Lin2001} 
and RPA-based corrections \cite{Chiesa2006} to minimize finite-size 
effects were not found essential for the RDF calculation. 

For the fully polarized and unpolarized fluid, we explored a range of
densities and of the parameter $\mu$. This required simulating several
different systems. We also needed to evaluate and extrapolate out, for
representative cases, time-step errors, population control bias, and
size effects. We plan to explore intermediate polarizations in a
future work.  

In the study we use two kinds of Jastrow-correlation-factors, one
better for the near-Jellium systems and one better for the near-ideal
systems.   

An important component of a computer experiment of a system of many
particles, a fluid, is the determination of suitable {\sl estimators}
to measure, through a statistical average, a given physical quantity,
an observable. Whereas the average from different estimators must give
the same result, the variance, the square of the statistical error,
can be different for different estimators. We compare the measure of
the histogram estimator for the RDF with a particular HFM one.

In ground state Monte Carlo simulations \cite{McMillan1965,Kalos1974},
unlike classical Monte Carlo 
simulations \cite{Hockney,Allen-Tildesley,Frenkel-Smit} and path
integral Monte Carlo simulations \cite{Ceperley1995}, one has to 
resort to the use of a trial wave function \cite{McMillan1965},
$\Psi$. While this is not a source of error, {\sl bias}, in diffusion
Monte Carlo simulation \cite{Kalos1974} of a system 
of Bosons, it is for a system of Fermions, due to the sign problem
\cite{Ceperley1991}. Another source of bias inevitably present in all 
three experiments is the finite size error.  

In a ground state Monte Carlo simulation, the energy has the
{\sl zero-variance} principle \cite{Ceperley1979}: as the trial wave
function approaches the exact ground state, the statistical error
vanishes. In a diffusion Monte Carlo simulation of a system of Bosons
the local energy of the trial wave function,
$E_L(\RR)=[H\Psi(\RR)]/\Psi(\RR)$, where $\RR$ denotes a configuration
of the system of particles and $H$ is the Hamiltonian, which we will here
assume to be real, is an unbiased estimator for the ground state. For
Fermions the ground state energy measurement is biased by the sign
problem. For observables $O$ which do not commute with the Hamiltonian
the local estimator $O_L(\RR)=[O\Psi(\RR)]/\Psi(\RR)$, is inevitably
biased by the choice of the 
trial wave function. A way to remedy to this bias can be the use of
the forward walking method \cite{Kalos1974b,Barnett1991} or the
reptation quantum Monte Carlo method \cite{Baroni1999}, to reach pure
estimates. Otherwise this bias can be made of leading order
$\delta^2$ with $\delta=\phi_0-\Psi$, where $\phi_0$ is the ground
state wave function, introducing the extrapolated measure
$\overline{O}^\text{ext}=2\langle O_L\rangle_f-\langle
O_L\rangle_{f_\text{vmc}}$, 
where the first statistical average, the mixed measure, 
is over the diffusion Monte Carlo (DMC) stationary
probability distribution $f$ and the second,
the variational measure, over the variational Monte
Carlo (VMC) probability distribution $f_\text{vmc}$, which can also be
obtained as the stationary probability distribution of a DMC without
branching \cite{Umrigar1993}.

One may follow different routes to determine estimators as the {\sl
  direct} microscopic one, the {\sl virial} route through the use of
the virial theorem, or the {\sl thermodynamic} route through the use
of thermodynamic identities. This aspect of finding out different ways
of calculating quantum properties in some ways resembles experimental
physics. The theoretical concept may be perfectly well defined but it
is up to the ingenuity of the experimentalist to find the best way of 
doing the measurement. Even what is meant by ``best'' is subject to
debate. In an unbiased experiment the different routes to the same
observable must give the same average.  

In this work we propose to use the Hellmann and Feynman
theorem as a direct route for the determination of estimators in a
diffusion Monte Carlo simulation. Some attempts in this direction have
been tried before \cite{Assaraf2003,Gaudoin2007}. The novelty of our
approach is a different definition of the correction to the
variational measure, necessary in the diffusion experiment, respect to
Ref. \cite{Assaraf2003} and the fact that the bias stemming
from the sign problem does not exhaust all the bias due to the choice
of the trial wave function, respect to Ref. \cite{Gaudoin2007}. 

The work is organized as follows: in Sec. \ref{sec:model} we introduce
the fluid model; in Sec. \ref{sec:ewald} we describe the Ewald sums
technique to treat the long range pair-potential; in
Sec. \ref{sec:dmc} we describe the fixed-nodes Diffusion Monte Carlo (DMC)
method; in Sec. \ref{sec:expectation} we describe several different ways
to evaluate expectation values in a DMC calculation; in
Sec. \ref{sec:twf} we describe the choice of the trial wave-function;
in Sec. \ref{sec:RDF} we define the RDF and describe some of 
its exact properties; the numerical results for the RDF are presented
in Section \ref{sec:results}; Sec. \ref{sec:conclusions} is for final
remarks. 

\section{The model}
\label{sec:model}
The Jellium is an assembly of $N$ electrons of charge $e$ moving in a
neutralizing background. The average particle number 
density is $n=N/\Omega$, where $\Omega$ is the volume of the fluid. In
the volume $\Omega$ there is
a uniform neutralizing background with a charge density
$\rho_b=-en$. So that the total charge of the system is zero.

In this paper lengths will be given in units of $a=(4\pi
n/3)^{-1/3}$. Energies will be given in Rydbergs $\hbar^2/(2ma_0^2)$,
where $m$ is the electron mass and $a_0=\hbar^2/(me^2)$ is the Bohr
radius. 

In these units the Hamiltonian of Jellium is
\bq 
H&=&-\frac{1}{r_s^2}\sum_{i=1}^N\nablab_{\rr_i}^2+V(\RR)~,\\
V&=&\frac{1}{r_s}\left(2\sum_{i<j}\frac{1}{|\rr_i-\rr_j|}+
\sum_{i=1}^Nr_i^2+
v_0\right)~,
\eq
where $\RR=(\rr_1,\rr_2,\ldots,\rr_N)$ with $\rr_i$ the
coordinate of the $i$th electron, $r_s=a/a_0$, 
and $v_0$ a constant containing the self energy of the background.

The kinetic energy scales as $1/r_s^2$ and the potential
energy (particle-particle, particle-background, and
background-background interaction) scales 
as $1/r_s$, so for small $r_s$ (high electronic densities), the
kinetic energy dominates and the electrons behave like an ideal gas. In
the limit of large $r_s$, the potential energy dominates and the
electrons crystallize into a Wigner crystal
\cite{Wigner1934,Ceperley1980}. No 
liquid phase is realizable within this model as the pair-potential
has no attractive parts even though a superconducting state
\anew{\cite{Leggett1975}} may still be possible (see chapter 8.9 of
Ref. \onlinecite{Giuliani-Vignale}).  

\subsection{Modified long range pair-potential}
\label{sec:lr}

The fluid model studied in this work is obtained modifying the Jellium
by replacing the $1/r$ Coulomb potential between the electrons with
the following long range bare pair-potential
\cite{Paziani2006} 
\bq
v_\mu(r)=\frac{\mbox{erf}(\mu r)}{r}~,
\eq
whose Fourier transform is
\bq
\tilde{v}_\mu(k)=\frac{4\pi}{k^2}e^{-\frac{k^2}{4\mu^2}}~.
\eq
When $\mu\to\infty$, we recover the standard Jellium model; in the
opposite limit $\mu\to 0$, we recover the non-interacting electron
gas. Notice that $v_\mu$ is a long range pair-potential with a
penetrable core, $v_\mu(0)=2\mu/\sqrt{\pi}$. So $\mu$ controls the
penetrability of two particles. For this kind of system it is lacking a
detailed study of the RDF. In this work we will only be concerned
about the fluid phase.

\section{Ewald sums}
\label{sec:ewald}
Periodic boundary conditions are necessary for extrapolating results
of the finite system to the thermodynamic limit. Suppose the bare 
pair-potential, in infinite space, is $v(r)$,
\bq
v(r)=\int\frac{d\kk}{(2\pi)^3}\,e^{-i\kk\cdot\rr}
\tilde{v}(k)~,~~~\tilde{v}(k)=\int d\rr\,e^{i\kk\cdot\rr}v(r)~.
\eq
The {\sl best} pair-potential of the finite system is given by the
image potential
\bq \label{imagep}
v_I(r)=\sum_{\LL}v(|\rr+\LL|)-\tilde{v}(0)/\Omega~.
\eq 
where the $\LL$ sum is over the Bravais lattice of the simulation cell
$\LL=(m_xL,m_yL,m_zL)$ where $m_x,m_y,m_z$ range over all
positive and negative integers and $\Omega=L^3$. We have also added a
uniform background of the same density but opposite charge. Converting
this to $k$-space and using the Poisson sum formula we get
\bq \label{poisson}
v_I(r)=\frac{1}{\Omega}\sum_{\kk}'\tilde{v}(k) e^{-i\kk\cdot\rr}~,
\eq
where the prime indicates that we omit the $\kk=0$ term; it cancels
out with the background. The $\kk$ sum is over reciprocal lattice
vectors of the simulation box $\kk_\nn=(2\pi n_x/L,$ $2\pi n_y/L,$ $2\pi
n_z/L)$  where $n_x,n_y,n_z$ range over all positive and negative
integers.  

Because both sums, Eq. (\ref{imagep}) and Eq. (\ref{poisson}), are so poorly
convergent \cite{Allen-Tildesley} 
we follow the scheme put forward by Natoli {\sl et al.}
\cite{Natoli1995} for approximating the image potential by a sum in 
$k$-space and a sum in $r$-space,
\bq
v_a(\rr)=\sum_\LL v_s(|\rr+\LL|)+\sum_{|\kk|\le
  k_c}v_l(k)e^{i\kk\cdot\rr}-\tilde{v}(0)/\Omega~,
\eq
where $v_s(r)$ is chosen to vanish smoothly as $r$ approaches $r_c$,
where $r_c$ is less than half of the distance across the simulation
box in any direction. If either $r_c$ or $k_c$ go to infinity then
$v_a\to v_I$. Natoli {\sl et al.} show that in order to minimize the
error in the potential, it is appropriate to minimize
$\chi^2=\int_\Omega[v_I(r)-v_a(r)]^2\,d\rr/\Omega$. And choose for
$v_s(r)$ an expansion in a fixed number of radial functions. This
same technique has also been applied to treat the Jastrow-correlation-factor
described in section \ref{sec:pseudo}. 

Now let us work with $N$ particles of charge $e$ in a periodic box
and let us compute the total potential energy of the unit
cell. Particles $i$ and $j$ are assumed to interact with a potential
$e^2v(r_{ij})=e^2v(|\rr_i-\rr_j|)$. The potential energy for the $N$
particle system is  
\bq
V=\sum_{i<j}e^2v_I(r_{ij})+\sum_{i}e^2v_M~,
\eq
where $v_M=\frac{1}{2}\lim_{r\to 0}[v_I(r)-v(r)]$ is the interaction
of a particle with its own images; it 
is a Madelung constant \anew{\cite{March-Tosi}} for particle $i$ interacting
with the perfect 
lattice of the simulation cell. If this term were not present,
particle $i$ would only see $N-1$ particles in the surrounding cells
instead of $N$.

\section{The fixed-nodes diffusion Monte Carlo (DMC) method}
\label{sec:dmc}
Consider the Schr\"odinger equation for the
many-body wave-function, $\phi(\RR,t)$ (the
wave-function can be assumed to be real, since both the real and
imaginary parts of the wave-function separately satisfy the
Schr\"odinger equation), in imaginary time, with a constant
shift $E_T$ in the zero of the energy. This is a diffusion equation in
a $3N$-dimensional 
space \cite{Anderson1976}. If $E_T$ is adjusted to be the 
ground-state energy, $E_0$, the asymptotic solution is a steady state 
solution, corresponding to the ground-state eigenfunction
$\phi_0(\RR)$ (provided $\phi(\RR,0)$ is not orthogonal to $\phi_0$).

Solving this equation by a random-walk process with branching is
inefficient, because the branching rate, which is proportional to the
total potential $V(\RR)$, can diverge to $+\infty$. This leads to
large fluctuations in the weights of the diffusers and to slow convergence
when calculating averages. However, the fluctuations, and hence the
statistical uncertainties, can be greatly reduced \cite{Kalos1974} by
the technique of importance sampling \cite{Hammersley}.

One simply multiplies the Schr\"odinger equation by a known trial
wave-function $\Psi(\RR)$ that approximate the unknown ground-state
wave-function, and rewrites it in terms of a new probability 
distribution
\bq
f(\RR,t)=\phi(\RR,t)\Psi(\RR)~,
\eq
whose normalization is given in Eq. (\ref{fnorm}). This leads to the
following diffusion equation 
\bq \label{is}
-\frac{\partial f(\RR,t)}{\partial t}=
-\lambda\nablab^2 f(\RR,t)+[E_L(\RR)-E_T]f(\RR,t)+
\lambda\nablab\cdot[f(\RR,t)\mathbf{F}(\RR)]~.
\eq
Here $\lambda=\hbar^2/(2m)$, $t$ is the imaginary time measured in
units of $\hbar$, $E_L(\RR)=[H\Psi(\RR)]/\Psi(\RR)$ is the local
energy of the trial wave-function, and
\bq
\mathbf{F}(\RR)=\nablab\ln\Psi^2(\RR)~.
\eq
The three terms on the right hand side of Eq. (\ref{is})
correspond, from left to right, to diffusion, branching, and drifting,
respectively. 

At sufficiently long times the solution to Eq. (\ref{is}) is
\bq \label{f-equilibrium}
f(\RR,t)\approx N_0\Psi(\RR)\phi_0(\RR)\exp[-(E_0-E_T)t]~,
\eq
where $N_0=\int\phi_0(\RR)\phi(\RR,0)\,d\RR$. If $E_T$ is adjusted to
be $E_0$, the asymptotic solution is a stationary solution and the
average $\langle E_L(\RR)\rangle_{f}$ of the local energy over the
stationary distribution gives the ground-state energy $E_0$. 
If we set the branching to zero $E_L(\RR)=E_T$ then this
average would be equal to the expectation value $\int
\Psi(\RR)H\Psi(\RR)\,d\RR$, since the stationary solution to
Eq. (\ref{is}) would then be $f=f_\text{vmc}=\Psi^2$. In other words,
without branching we would  
obtain the variational energy of $\Psi$, rather than $E_0$, as in a
Variational Monte Carlo (VMC) calculation.

The time evolution of $f(\RR,t)$ is given by
\bq
f(\RR^{'},t+\tau)=\int d\RR G(\RR^{'},\RR;\tau)f(\RR,t)~,
\eq
where the Green's function
$G(\RR^{'},\RR;\tau)=\Psi(\RR^\prime)\langle\RR^\prime|
\exp[-\tau(H-E_T)]|\RR\rangle\Psi^{-1}(\RR)$ is a transition
probability for moving  
the set of coordinates from $\RR$ to $\RR^{'}$ in a time $\tau$. Thus
$G$ is a solution of the same differential equation, Eq. (\ref{is}), but
with the initial condition $G(\RR^{'},\RR;0)=\delta(\RR^{'}-\RR)$.
For short times $\tau$ an approximate solution for $G$ is
\bq \label{Green}
G(\RR^{'},\RR;\tau)=(4\pi\lambda\tau)^{-3N/2}
e^{-|\RR^{\prime}-\RR-\lambda\tau \mathbf{F}(\RR)|^2/4\lambda\tau}
e^{-\tau\{[E_L(\RR)+E_L(\RR^{\prime})]/2-E_T\}}+O(\tau^2)~.
\eq
To compute the ground-state energy and other expectation values, the
$N$-particle distribution function $f(\RR,t)$ is represented, in
diffusion Monte Carlo, by an average over a time series of
generations of walkers each of which consists of a fixed number
of $n_w$ walkers. A walker is a pair $(\RR_\alpha,\omega_\alpha)$,
$\alpha=1,2,\ldots,n_w$, with $\RR_\alpha$ a $3N$-dimensional particle
configuration with statistical weight $\omega_\alpha$. At time $t$,
the walkers represent a random realization of the $N$-particle
distribution,
$f(\RR,t)=\sum_{\alpha=1}^{n_w}\omega^t_\alpha\delta(\RR-\RR^t_\alpha)$.
The ensemble is initialized with a VMC sample from
$f(\RR,0)=\Psi^2(\RR)$, with $\omega^0_\alpha=1/n_w$ for all $\alpha$. 
Note that if the trial wave-function were the exact ground-state then
there would be no branching and it would be sufficient $n_w=1$. 
A given walker $(\RR^t,\omega^t)$ is advanced in time
(diffusion and drift) as
$\RR^{t+\tau}=\RR^t+\chi+\lambda\tau\nablab\ln\Psi^2(\RR^t)$ 
where $\chi$ is a normally distributed random $3N$-dimensional vector
with variance $2\lambda\tau$ and zero mean
\anew{\cite{Kalos-Whitlock}}. In order to satisfy 
detailed balance we accept the move with a probability
$A(\RR,\RR^\prime;\tau)=\min[1,W(\RR,\RR^\prime)]$, where 
$W(\RR,\RR^\prime)=[G(\RR,\RR^\prime;\tau)\Psi^2(\RR^\prime)]/[
G(\RR^\prime,\RR;\tau)\Psi^2(\RR)]$. This step would be unnecessary if
$G$ were the exact Green's function, since $W$ would be unity.
Finally, the weight $\omega^t_\alpha$ is replaced by
$\omega^{t+\tau}_\alpha=\omega^t_\alpha\Delta\omega^t_\alpha$
(branching), with  
$\Delta\omega^t_\alpha=\exp\{-\tau[(E_L(\RR^t_\alpha)
  +E_L(\RR^{t+\tau}_\alpha))/2-E_T]\}$.  

However, for the diffusion interpretation to be valid, $f$ must always
be positive, since it is a probability distribution. But we know that
the many-fermions wave-function $\phi(\RR,t)$, being antisymmetric
under exchange of a pair of particles of the parallel spins, must have
nodes, {\sl i.e.} points $\RR$ where it vanishes. In the fixed-nodes
approximation one restricts the diffusion process to walkers that do
not change the sign of the trial wave-function. One can easily
demonstrate that the resulting energy, $\langle E_L(\RR)\rangle_{f}$, will
be an upper bound to the exact ground-state 
energy; the best possible upper bound with the given boundary
condition \cite{Ceperley1991}.

A detailed description of the algorithm used for the DMC calculation
can be found in Ref. \onlinecite{Umrigar1993}.

\section{Expectation values in DMC}
\label{sec:expectation}

In a DMC calculation there are various different possibilities to
measure the expectation value of a physical observable, as for
example the RDF. If $\langle {\cal O}\rangle_f$ is the measure and
$\langle\ldots\rangle_f$ the statistical average over the 
probability distribution $f$ we will, in the following, use the word
{\sl estimator} to indicate the function ${\cal O}$ itself, unlike the
more common use of the word to indicate the usual Monte Carlo
estimator $\sum_{i=1}^{\cal N}{\cal O}_i/{\cal N}$ of the average,
where $\{{\cal O}_i\}$ is the set obtained evaluating ${\cal O}$ over a
finite number ${\cal N}$ of points distributed according to
$f$. Whereas the average from different estimators must give the same
result, the variance, the square of the statistical error, can be
different for different estimators. 

\subsubsection{The local estimator and the extrapolated measure}

To obtain ground-state expectation values of quantities $O$ that do
not commute with the Hamiltonian we introduce the local estimator
$O_L(\RR)=[O\Psi(\RR)]/\Psi(\RR)$ and then compute the average over
the DMC walk, the so called mixed measure,
$\overline{O}^\text{mix}=\langle 
O_L(\RR)\rangle_{f}$ $=$ $\int\phi_0(\RR)O\Psi(\RR)\,d\RR/$ 
$\int\phi_0(\RR)\Psi(\RR)\,d\RR$. This is inevitably biased by the
choice of the trial wave-function. A way to remedy to this bias is the
use of the forward walking method \cite{Kalos1974b,Barnett1991} or the
reptation quantum Monte Carlo method \cite{Baroni1999} to reach pure
estimates. Otherwise this bias can be made of leading order
$\delta^2$, with $\delta=\phi_0-\Psi$, introducing the extrapolated
measure 
\bq \label{extra1}
\overline{O}^\text{ext}=2\overline{O}^{mix}-\overline{O}^{var}~,
\eq
where $\overline{O}^{var}=\langle O_L\rangle_{f_\text{vmc}}$ is the
variational measure. If the mixed measure equals the variational
measure then the trial wave-function has maximum overlap with the
ground-state. 

\subsubsection{The Hellmann and Feynman measure}
\label{sec:ZVZB}

Toulouse {\sl et al.} \cite{Toulouse2007,Assaraf2003} observed that the
{\sl zero-variance} property of the energy \cite{Ceperley1979} can be
extended to an arbitrary observable, $O$, by expressing it as an
energy derivative through the use of the Hellmann-Feynman theorem.

In a DMC calculation the Hellmann-Feynman theorem takes a form
different from the one in a VMC calculation. Namely we start with the
eigenvalue expression $(H^\lambda-E^\lambda)\Psi^\lambda=0$ for the
ground-state of the perturbed Hamiltonian $H^\lambda=H+\lambda O$,
take the derivative with respect to $\lambda$, multiply on the right
by the ground-state at $\lambda=0$, $\phi_0$, and integrate
over the particle coordinates to get 
\bq
\int d\RR\,\phi_0(H^\lambda-E^\lambda)\frac{\partial\Psi^\lambda}
{\partial\lambda}=\int d\RR\,\phi_0
\left(\frac{\partial E^\lambda}{\partial\lambda}-
\frac{\partial H^\lambda}{\partial\lambda}\right)\Psi^\lambda~.
\eq 
Then we notice that due to the Hermiticity of the Hamiltonian, at
$\lambda=0$ the left hand side vanishes, so that we get \cite{Fantoni13}
\bq
\left.\frac{\int d\RR\,\phi_0 O\Psi^\lambda}{\int
  d\RR\,\phi_0\Psi^\lambda}\right|_{\lambda=0}=
\left.\frac{\partial E^\lambda}{\partial\lambda}\right|_{\lambda=0}~.
\eq
This relation holds only in the $\lambda\to 0$ limit
unlike the more common form \cite{LandauQM} which holds
for any $\lambda$. Also it resembles Eq. (3) of
Ref. \onlinecite{Gaudoin2007}.  

Given $E^\lambda=\int d\RR \phi_0(\RR)H^\lambda\Psi^\lambda(\RR)/\int
d\RR\phi_0(\RR)\Psi^\lambda(\RR)$ the ``Hellmann and Feynman'' (HFM)
measure in a DMC calculation is 
\bq \label{zvzb-d}
\overline{O}^{HFM}=\left.\frac{dE^\lambda}{d\lambda}\right|_{\lambda=0}\approx
\langle O_L(\RR)\rangle_{f}+\langle\Delta
O_L^{\alpha}(\RR)\rangle_{f}+\langle\Delta O_L^{\beta}(\RR)\rangle_{f}~.
\eq 
The $\alpha$ correction is \cite{Fantoni13}
\bq \label{zv-d}
\Delta O_L^{\alpha}(\RR)=\left[\frac{H\Psi^\prime}{\Psi^\prime}-E_L(\RR)\right]
\frac{\Psi^\prime(\RR)}{\Psi(\RR)}~.
\eq
This expression coincides with Eq. (18) of
Ref. \onlinecite{Toulouse2007}. In a VMC calculation  
this term, usually, does not contribute to the average, with respect to
$f_\text{vmc}=\Psi^2$, due to the Hermiticity of the
Hamiltonian. This is of course not true in a DMC calculation.
We will then define a Hellmann and Feynman variational (HFMv)
estimator as $O^{HFMv}=O_L(\RR)+\Delta O_L^{\alpha}(\RR)$. The $\beta$
correction is \cite{Fantoni13}  
\bq \label{zb-d}
\Delta O_L^{\beta}(\RR)=[E_L(\RR)-E_0]\frac{\Psi^\prime(\RR)}{\Psi(\RR)}~,
\eq
where $E_0=E^{\lambda=0}$. Which differs from Eq. (19) of
Ref. \onlinecite{Toulouse2007} by a factor of one half. 
This term is necessary in a DMC calculation not to bias the
measure. The extrapolated Hellmann and Feynman measure will then be
\bq \label{zvzb-e}
\overline{O}^\text{HFM-ext}=2\overline{O}^\text{HFM}-\langle
O^\text{HFMv}\rangle_{f_\text{vmc}}~.
\eq 
Both corrections $\alpha$ and $\beta$ to the local 
estimator depends on the {\sl auxiliary} function,
$\Psi^\prime=\partial\Psi^\lambda/\partial\lambda|_{\lambda=0}$.
Of course if we had chosen $\Psi^{\lambda=0}$, on the left hand side
of Eq. (\ref{zvzb-d}), as the exact ground state wave-function, $\phi_0$,
instead of the trial wave-function, then both corrections would have
vanished. When the trial wave-function is sufficiently close to the
exact ground state function a good approximation to the auxiliary
function can be obtained from first order perturbation theory for
$\lambda\ll 1$. So the Hellmann and Feynman measure is affected by the
new source of bias due to the choice of the auxiliary function
independent from the bias due to the choice of the trial wave-function.  

It is convenient to rewrite Eqs. (\ref{zv-d}) and (\ref{zb-d}) in
terms of the logarithmic derivative
$Q(\RR)=\Psi^\prime(\RR)/\Psi(\RR)$ as follows
\bq \label{zvq}
\Delta O_L^{\alpha}(\RR)&=&-\frac{1}{r_s^2}\sum_{k=1}^N[\nablab_{\rr_k}^2Q(\RR)+
2\vv_k(\RR)\cdot\nablab_{\rr_k}Q(\RR)]~,\\ \label{zbq}
\Delta O_L^{\beta}(\RR)&=&[E_L(\RR)-E]Q(\RR)~,
\eq
where $\vv_k(\RR)=\nablab_{\rr_k}\ln\Psi(\RR)$ is the drift
velocity of the trial wave-function. For each observable a specific
form of $Q$ has to be chosen.

\section{Trial wave-function}
\label{sec:twf}

We chose the trial wave-function of the Bijl-Dingle-Jastrow
\cite{Bijl1940} or product form 
\bq \label{twf}
\Psi(\RR)\propto D(\RR)\exp\left(-\sum_{i<j}u(r_{ij})\right)~.
\eq

The function $D(\RR)$ is the exact wave-function of the non-interacting
fermions (the Slater determinant) and serves to give the trial
wave-function the desired antisymmetry 
\bq
D(\RR)=\frac{1}{\sqrt{N_+!}}\det(\varphi^+_{n,m})
\frac{1}{\sqrt{N_-!}}\det(\varphi^-_{n,m})~,
\eq
where for the fluid phase
$\varphi^\sigma_{n,m}=e^{i\kk_n\cdot\rr_m}\delta_{\sigma_m,\sigma}/\sqrt{\Omega}$
with $\kk_n$ a reciprocal lattice vector of the simulation box such that 
$|\kk_n|\le k^\sigma_F$, $\sigma$ the $z$-component of the spin ($\pm
1/2$), $\rr_m$ the coordinates of particle $m$, and $\sigma_m$ its
spin $z$-component. For the unpolarized fluid there are two separate 
determinants for the spin-up and the spin-down states because the
Hamiltonian is spin independent. For the polarized fluid there is a
single determinant. For the general case of $N_+$ spin-up particles
the polarization will be $\zeta=(N_+-N_-)/N$ and the 
Fermi wave-vector for the spin-up (spin-down) particles will be
$k_F^\pm=(1\pm\zeta)^{1/3}k_F$ with $k_F=(3\pi^2
n)^{1/3}=(9\pi/4)^{1/3}/(a_0r_s)$ the Fermi wave-vector of the
paramagnetic fluid. 
On the computer we fill closed shells so that $N_\sigma$ is always
odd. We only store $\kk_n$ for each pair $(\kk_n,-\kk_n)$ and use
sines and cosines instead of $\exp(i\kk_n\cdot\rr_i)$ and
$\exp(-i\kk_n\cdot\rr_j)$. 

The second factor (the Jastrow factor) includes in an approximate way
the effects of particle correlations, through the
``Jastrow-correlation-factor'', $u(r)$, which is repulsive. 

\subsection{The Jastrow-correlation-factor}
\label{sec:pseudo}
Neglecting the cross term between the Jastrow and the Slater
determinant in Eq. (\ref{jel}) (third term) and the Madelung constant,
the variational energy per particle can be approximated as follows, 
\bq \nonumber
e_V&=&\frac{\langle E_L(\RR) \rangle_{f}}{N}
=\frac{\int\Psi(\RR)H\Psi(\RR)\,d\RR}
{N} \approx e_F+\frac{1}{2\Omega}
\sum^\prime_{\kk}[e^2\tilde{v}_\mu(k)-2\lambda k^2\tilde{u}(k)][S(k)-1]+\\
&&\frac{1}{N\Omega^2}\sum^\prime_{\kk,\kk^\prime}\lambda
\kk\cdot\kk^\prime\tilde{u}(k)\tilde{u}(k^\prime)\langle
\rho_{\kk+\kk^\prime}\rho_{-\kk}\rho_{-\kk^\prime}\rangle_{f}+\ldots~,
\eq
where $e_F=(3/5)\lambda \sum_\sigma N_\sigma(k_F^\sigma)^2/N$ is the
non-interacting fermions energy per particle, $\tilde{u}(k)$ is the
Fourier transform of the Jastrow-correlation-factor $u(r)$,
$\tilde{v}_\mu(k)=4\pi\exp(-k^2/4\mu^2)/k^2$  
is the Fourier transform of the bare pair-potential, $S(k)$ is the 
static structure factor for a given $u(r)$ (see Sec. \ref{sec:ssf}),
$\rho_\kk=\sum_{i=1}^N\exp(i\kk\cdot\rr_i)$ is the Fourier transform of
the total number density $\rho(\rr)=\sum_i\delta(\rr-\rr_i)$, and the
trailing dots stand for the additional terms coming from the exclusion
of the $j=k$ term in the last term of Eq. (\ref{jel}). Next we
make the Random Phase Approximation \cite{Feynman} and we keep only
the terms with $\kk+\kk^\prime=\mathbf{0}$ in the last term. This gives
\bq
e_V\approx e_F+\frac{1}{2\Omega}
\sum^\prime_{\kk}\Big\{[e^2\tilde{v}_\mu(k)-2\lambda
k^2\tilde{u}(k)][S(k)-1]-2n\lambda[k\tilde{u}(k)]^2S(k)\Big\}+\ldots~.
\eq
In the limit $k\to 0$ we have to cancel the Coulomb singularity and we
get $\tilde{u}^2(k)=me^2\tilde{v}_\mu(k)/(\hbar^2 nk^2)\simeq
[(4\pi e^2/k^2)/(\hbar\omega_p)]^2$ (where $\omega_p=\sqrt{4\pi
ne^2/m}$ is the plasmon frequency) or in adimensional units
\bq \label{urpa}
\tilde{u}(k)=\sqrt{\frac{r_s}{3}}\frac{4\pi}{k^2}~,~~~\text{small $k$}~.
\eq
This determines the correct behavior of $\tilde{u}(k)$ as $k\to 0$ or
the long range behavior of $u(r)$
\bq
u(r)=\sqrt{\frac{r_s}{3}}\frac{1}{r}~,~~~\text{large $r$}~.
\eq

Now to construct the approximate Jastrow-correlation-factor, we start from the
expression 
\bq
\epsilon = e_F+\frac{1}{2\Omega}
\sum^\prime_{\kk}[e^2\tilde{v}_\mu(k)-{\cal A}\lambda
k^2\tilde{u}(k)][S(k)-1]~,
\eq
and use the following perturbation approximation, for how $S(k)$
depends on $\tilde{u}(k)$ \cite{Gaskell61,*Gaskell62}, 
\bq \label{crpa}
\frac{1}{S(k)}=\frac{1}{S^x(k)}+{\cal B} n\tilde{u}(k)~,
\eq
where ${\cal A}$ and ${\cal B}$ are constant to be determined and $S^x(k)$
the structure factor for the non-interacting fermions 
(see Eq. (\ref{Sx2})), which is $S^x=\sum_\sigma S^x_{\sigma,\sigma}$ with
\bq
S^x_{\sigma,\sigma}(k)=\left\{
\begin{array}{ll} 
\displaystyle
\frac{n_\sigma}{n}\frac{y_\sigma}{2}(3-y_\sigma^2) & y_\sigma<1\\
\displaystyle
\frac{n_\sigma}{n} & \text{else}
\end{array}
\right.
\eq
where $n_{\sigma}=N_\sigma/\Omega$ and $y_{\sigma}=k/(2k_F^\sigma)$.

Minimizing $\epsilon$ with respect to $u(k)$, we obtain
\cite{Ceperley2004} 
\bq \label{jastrowrpam}
{\cal B} n\tilde{u}(k)=-\frac{1}{S^x(k)}+\left[\frac{1}{S^x(k)}+
\frac{{\cal B} ne^2\tilde{v}_\mu(k)}{\lambda {\cal A} k^2}\right]^{1/2}~,
\eq
This form is optimal at both long and short distances but
not necessarily in between. In particular, for any value of 
$\zeta$, the small $k$ behavior of
$\tilde{u}(k)$ is $\sqrt{2r_s/3{\cal A}{\cal B}}(4\pi/k^2)$ which means that
\bq \label{jastrowlr}
u(r)=\sqrt{\frac{2r_s}{3{\cal A}{\cal B}}}\frac{1}{r}~,~~~
\text{large $r$}~.
\eq
The large $k$ behavior of $\tilde{u}(k)$ is
$(r_s/{\cal A})\tilde{v}_\mu(k)/k^2$, for any value of $\zeta$, which in
$r$ space translates into
\bq \label{jastrowsr}
\left.\frac{du(r)}{dr}\right|_{r=0}=\left\{
\begin{array}{ll}
\displaystyle
-\frac{r_s}{2{\cal A}}& \mu\to \infty\\
\displaystyle
0& \mbox{$\mu$ finite}
\end{array}
\right.
\eq
In order to satisfy the cusp condition for particles of antiparallel
spins 
(any reasonable Jastrow-correlation-factor has to obey to the cusp conditions
(see Ref. \onlinecite{Foulkes2001} Section IVF) which prevent the local 
energy from diverging whenever any two electrons ($\mu=\infty$) come
together) we need to choose ${\cal A}=1$, then the correct behavior at
large $r$ (\ref{urpa}) is obtained  fixing ${\cal B}=2$
(see Note \onlinecite{rpanote}). We will call this Jastrow ${\cal J}_1$ in the
following. 

It turns out that, at small $\mu$, but not for the Coulomb case, a better
choice is given by \cite{Ceperley78}
\bq \label{jastrowrpa}
2n\tilde{u}(k)=-\frac{1}{S^x(k)}+\left[\left(\frac{1}{S^x(k)}\right)^2+
\frac{2ne^2\tilde{v}_\mu(k)}{\lambda k^2}\right]^{1/2}~,
\eq
which still has the correct long (\ref{jastrowlr}) and short
(\ref{jastrowsr}) range behaviors. We will call this Jastrow ${\cal J}_2$ in the
following. This is expected since, differently from ${\cal J}_1$, ${\cal J}_2$ satisfies
the additional exact requirement $\lim_{\mu\to 0}u(r)=0$, as 
immediately follows from the definition (\ref{jastrowrpa}). Then, as
confirmed by our results (see Sec. \ref{sec:jastrow-results})), at small
$\mu$ (and any $r_s$), the trial wave-function is expected to be very
close to the stationary solution of the diffusion problem.

\subsection{The backflow and three-body correlations}
\label{sec:bf+3b}
As shown in Appendix \ref{app:bf+3bd}, the trial wave-function of
Eq. (\ref{twf}) can be further improved by adding three-body (3B) and
backflow (BF) correlations \cite{Kwon1993,Kwon1998} as follows 
\bq \label{twfi}
\Psi(\RR)=\tilde{D}(\RR)\exp\left[-\sum_{i<j}
\tilde{u}(r_{ij})-\sum_{l=1}^N\mathbf{G}(l)\cdot\mathbf{G}(l)\right]~.
\eq
Here
\bq
\tilde{D}(\RR)=\frac{1}{\sqrt{N_+!}}\det(\tilde{\varphi}^+_{n,m})
\frac{1}{\sqrt{N_-!}}\det(\tilde{\varphi}^-_{n,m})~,
\eq
with $\tilde{\varphi}^\sigma_{n,m}=e^{i\kk_n\cdot\xx_m}\delta_{\sigma_m,\sigma}/
\sqrt{\Omega}$ and $\xx_m$ quasi-particle coordinates defined as 
\bq
\xx_i=\rr_i+\sum_{j\neq i}^N\eta(r_{ij})(\rr_i-\rr_j)~.
\eq
The displacement of the quasi-particle coordinates $\xx_i$ from the
real coordinate $\rr_i$ incorporates effects of hydrodynamic backflow 
\cite{Feynman1956}, and changes the nodes of the trial
wave-function. The backflow correlation function $\eta(r)$, is
parametrized as \cite{Kwon1998}
\bq
\eta(r)=\lambda_B\frac{1+s_Br}{r_B+w_Br+r^4}~,
\eq
which has the long-range behavior $\sim 1/r^3$.

Three-body correlations are included through the vector functions
\bq
\mathbf{G}(i)=\sum_{j\neq i}^N\xi(r_{ij})(\rr_i-\rr_j)~.
\eq
We call $\xi(r)$ the three-body correlation function which is
parametrized as \cite{Panoff1989}
\bq
\xi(r)=a\exp\left\{-[(r-b)c]^2\right\}~.
\eq
To cancel the two-body term arising from
$\mathbf{G}(l)\cdot\mathbf{G}(l)$, we use 
$\tilde{u}(r)=u(r)-2\xi^2(r)r^2$

The backflow and three-body correlation functions are then chosen to
decay to zero with a zero first derivative at the edge of the
simulation box.

\subsection{Optimized parameters}
\label{sec:opt}
 
Optimizing the trial wave-function (see Ref. \onlinecite{Foulkes2001} Section
VII) is extremely important for a 
fixed-nodes DMC calculation as, even if the Jastrow-correlation-factor is
parameter free, the backflow changes the nodes. We carefully studied
how the RDF depends on the quality of the trial wave-function choosing
a simple Slater determinant (S) (Eq. (\ref{twf}) without the Jastrow
factor), a Slater-Jastrow (SJ) (Eq. (\ref{twf})), and a
Slater-Jastrow with the backflow and three-body corrections (SJ+BF+3B)
(Eq. (\ref{twfi})).

In Table \ref{tab:opt} we report the optimized parameters for the
backflow and three-body correlation functions for a system of $N=54$
and $\zeta=0$ at various $r_s$ and $\mu$. We have used these values of
the parameters in all subsequent calculations, unrespective of the
value of $\zeta$.  

\begin{table}
\caption{Optimized variational parameters of backflow and three-body
  correlation functions for $N=54$ and $\zeta=0$ and various
  combinations af $r_s$ and $\mu$.
\label{tab:opt}}
{\scriptsize
\begin{ruledtabular}
\begin{tabular}{ccccccccc}
$r_s$&$\mu$&$\lambda_B$&$s_B$&$r_B$&$w_B$&$a$&$b$&$c$\\
\hline
10&1/2&-&-&-&-&-&-&-\\
10&1&8.408d-4&1.658d+2&-1.383d-3&3.168&0.447&-0.212&1.036\\
10&2&7.189d-5&9.793d+2&9.478d-6&0.446&1.379d+1&-3.688&0.450\\
10&4&1.116d-4&6.522d+2&-2.553d-5&0.179&5.981d+1&-4.773&0.462\\
10&$\infty$&0.781&-0.499&0.324&2.958&0.514&0.327&1.358\\
\hline
5&1/2&-&-&-&-&-&-&-\\
5&1&-&-&-&-&-&-&-\\
5&2&2.768d-2&-0.420&0.893&-0.673&1.322d+6&-9.003&0.408\\
5&4&0.331&-0.680&1.467& 1.442&2.729d+1&-2.607&0.659\\
5&$\infty$&0.161&-0.585&0.335&0.841&0.802&-7.310d-2&1.344\\
\hline
2&1/2&-&-&-&-&-&-&-\\
2&1&-&-&-&-&-&-&-\\
2&2&-&-&-&-&-&-&-\\
2&4&5.272d-2&-1.616&1.732&1.687d-2&804.135&-2.875&0.847\\
2&$\infty$&5.018d-2&-1.221&0.393&0.681&1.655&-0.596&1.229\\
\hline
1&1/2&-&-&-&-&-&-&-\\
1&1&-&-&-&-&-&-&-\\
1&2&-&-&-&-&-&-&-\\
1&4&1.187d-2&-6.834&0.495&1.295&0.186&0.489&4.739\\
1&$\infty$&2.1945d-2&-3.086&0.320&1.631&0.306&0.367&2.467\\
\end{tabular}
\end{ruledtabular}
}
\end{table}

\begin{figure}[H]
\begin{center}
\includegraphics[width=8cm]{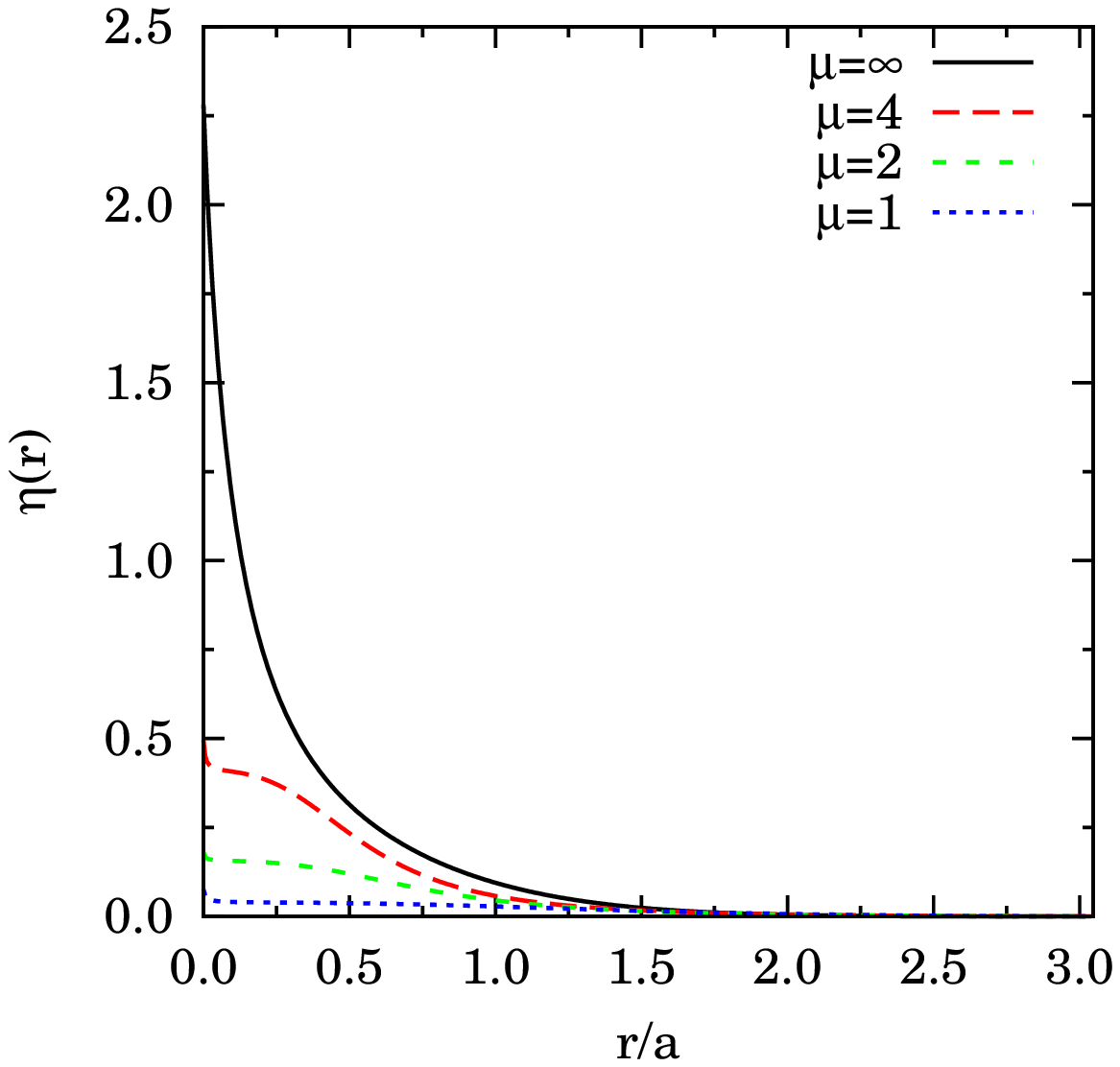}
\includegraphics[width=8cm]{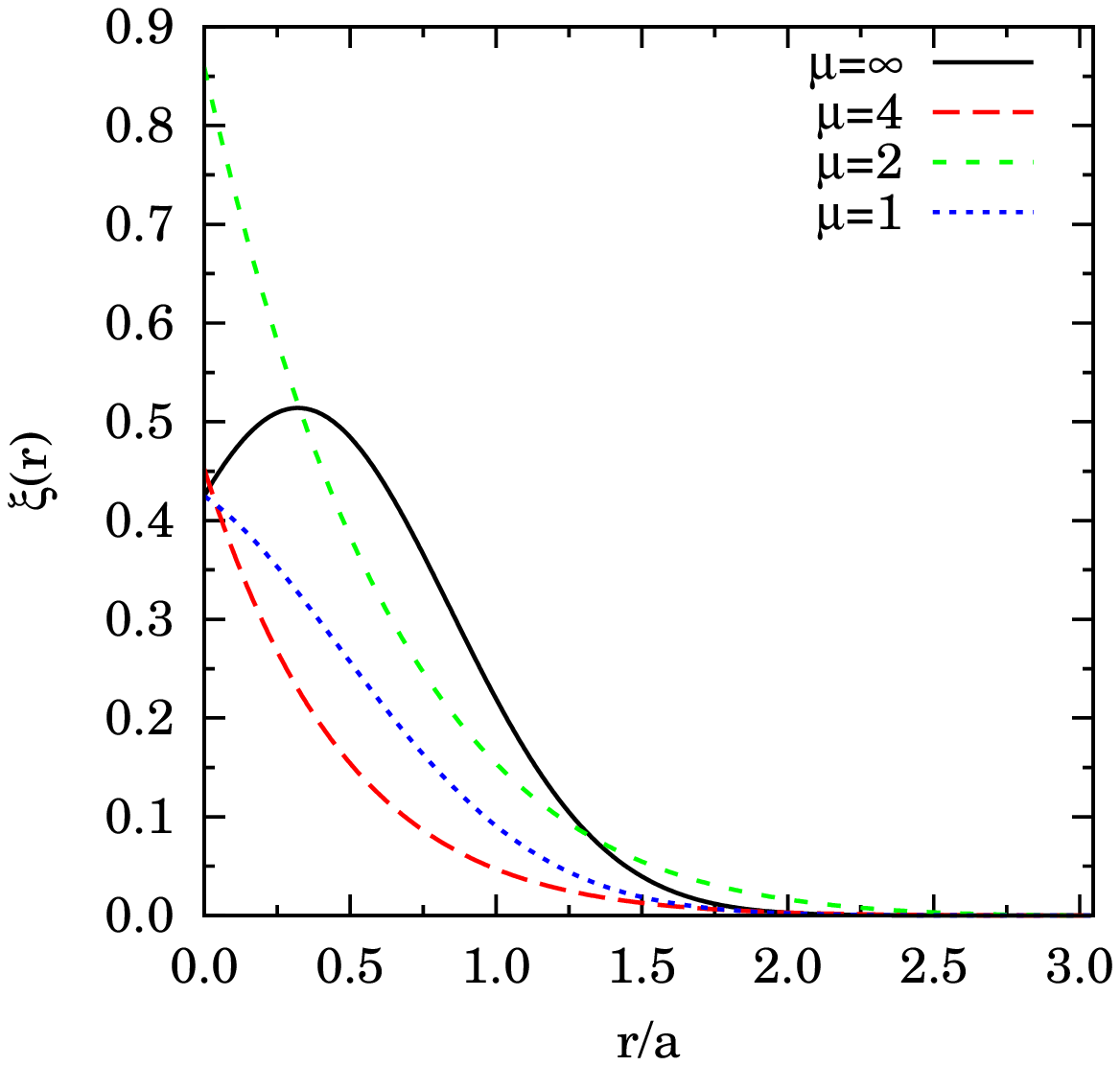}
\end{center}
\caption{Shows the optimized correlation functions $\eta$ and $\xi$ for
$N=54, \zeta=0,$ and $r_s=10$ and different values of $\mu$.}
\label{fig:opt10}
\end{figure}

In Fig. \ref{fig:opt10} we show the optimized $\eta$ and $\xi$ for
$N=54, \zeta=0, r_s=10$. The optimization of
the 7 parameter dependent trial wave-function gives a backflow
correlation $\eta$ ordered in $\mu$ but a three-body correlation $\xi$
erratic in $\mu$. As one moves away from the Coulomb $\mu\to\infty$
case the system of particles becomes less interacting and the
relevance of the backflow and three-body correlations diminishes. This
is supported by the fact that at $\mu=4,2,1$, in correspondnce of the
erratic behavior, the effect of the three-body correlations on the
expectation value of the energy is irrelevant.  

%

\section{The radial distribution function (RDF)}
\label{sec:RDF}

The main purpose of the present work is to determine the radial
distribution function (RDF) of our fluid model through the DMC
calculation. 

\subsubsection{Definition of the radial distribution function}

The spin-resolved RDF is defined as \anew{\cite{Hill,Feenberg}} 
\bq \label{RDF}
g_{\sigma,\sigma^{\prime}}(\rr,\rr^{'})&=&
\frac{\left\langle\sum_{i,j\neq
    i}\delta_{\sigma,\sigma_i}\delta_{\sigma^{\prime},\sigma_j}
\delta(\rr-\rr_i)\delta(\rr^{'}-\rr_j)\right\rangle}
{n_{\sigma}(\rr)n_{\sigma^{\prime}}(\rr^{'})}~,\\
n_{\sigma}(\rr)&=&\left\langle\sum_{i=1}^N\delta_{\sigma,\sigma_i}
\delta(\rr-\rr_i)\right\rangle~,
\eq
where here, and in the following, $\langle\ldots\rangle$ will denote
the expectation value respect to the ground-state. Two exact
conditions follow immediately from the definition:
i. the zero-moment sum rule
\bq \label{0msr}
\sum_{\sigma,\sigma^\prime}\int d\rr d\rr^\prime\,n_{\sigma}(\rr)
n_{\sigma}(\rr^\prime)[g_{\sigma,\sigma^{\prime}}(\rr,\rr^{'})-1]=-N~,
\eq
also known as the charge (monopole) sum rule in the sequence of
multipolar sum rules in the framework of charged fluids
\cite{Martin88}, ii. $g_{\sigma,\sigma}(\rr,\rr)=0$ due to the Pauli
exclusion principle.

For the homogeneous and isotropic fluid
$n_{\sigma}(\rr)=N_\sigma/\Omega$ where $N_\sigma$ is the number of
particles of spin $\sigma$ and $g_{\sigma,\sigma^\prime}$ depends 
only on the distance $r=|\rr-\rr^{'}|$, so that
\bq \label{RDFhi}
g_{\sigma,\sigma^{\prime}}(r)=\frac{1}{4\pi r^2}
\frac{\Omega}{N_\sigma N_{\sigma^{\prime}}}
\left\langle\sum_{i,j\neq i}\delta_{\sigma,\sigma_i}\delta_{\sigma^{\prime},\sigma_j} 
\delta(r-r_{ij})\right\rangle~.
\eq
The total (spin-summed) radial distribution function will be
\bq \nonumber
g(r)&=&\frac{1}{n^2}\sum_{\sigma,\sigma^\prime}n_\sigma n_{\sigma^\prime}
g_{\sigma,\sigma^{\prime}}(r)\\
&=&\left(\frac{1+\zeta}{2}\right)^2g_{+,+}(r)
+\left(\frac{1-\zeta}{2}\right)^2g_{-,-}(r)
+\frac{1-\zeta^2}{2}g_{+,-}(r)~.
\eq

\subsubsection{From the structure to the thermodynamics}

As it is well known the knowledge of the RDF gives access to the
thermodynamic properties of the system.
The mean potential energy per particle can be directly obtained from
$g(r)$ and the bare pair-potential $v_\mu(r)$ as follows
\bq \label{ep}
e_p=\sum_{\sigma,\sigma^\prime}\frac{n_\sigma n_{\sigma^\prime}}{2n}
\int d\rr\,e^2v_\mu(r)[g_{\sigma,\sigma^\prime}(r)-1]~,
\eq
where we have explicitly taken into account of the background
contribution. Suppose that $e_p(r_s)$ is known as a function of the
coupling strength $r_s$. The virial theorem for a system with Coulomb
interactions ($v_\infty(r)=1/r$) gives $N(2e_k+e_p)=3P\Omega$ with
$P=-d(Ne_0)/d\Omega$ 
the pressure and $e_0=e_k+e_p$ the mean total ground-state energy per
particle. We then find 
\bq
e_p(r_s)=2e_0(r_s)+r_s\frac{de_0(r_s)}{dr_s}=
\frac{1}{r_s}\frac{d}{dr_s}[r_s^2e_0(r_s)]~,
\eq
which integrates to
\bq
e_0(r_s)=e_F+\frac{1}{r_s^2}\int_0^{r_s}dr_s^\prime\,r_s^\prime
  e_p(r_s^\prime)~. 
\eq

We can rewrite the ground-state energy per particle of the ideal Fermi
gas, in reduced units, as
\bq
e_F=\left(\frac{9\pi}{4}\right)^{2/3}\frac{3}{10}\phi_5(\zeta)
\frac{1}{r_s^2}~, 
\eq
where $\phi_n(\zeta)=(1-\zeta)^{n/3}+(1+\zeta)^{n/3}$.
And for the exchange potential energy per particle in the Coulomb case 
\bq \label{epx}
e_p^x=-\left(\frac{2}{3\pi^5}\right)^{1/3}\frac{9\pi}{8}\phi_4(\zeta)
\frac{1}{r_s}~,
\eq
which follows from Eq. (\ref{ep}) and Eqs. (\ref{gx1})-(\ref{gx2}).
The expression for finite $\mu$ can be found in
Ref. \onlinecite{Paziani2006} (see their Eqs. (15)-(16)). 

\subsubsection{Definition of the static structure factor}
\label{sec:ssf}

If we introduce the microscopic spin dependent number density
\bq
\rho_{\sigma}(\rr)=\sum_{i=1}^N\delta_{\sigma,\sigma_i}\delta(\rr-\rr_i)~,
\eq
and its Fourier transform $\rho_{\kk,\sigma}$, then the spin-resolved
static structure factors are defined as
$S_{\sigma,\sigma^{\prime}}(\kk)=
\langle\rho_{\kk,\sigma}\rho_{-\kk,\sigma^{\prime}}\rangle/N$,
which, for the homogeneous and isotropic fluid, can be rewritten as  
\bq
S_{\sigma,\sigma^{\prime}}(k)=\frac{n_\sigma}{n}\delta_{\sigma,\sigma^\prime}+
\frac{n_\sigma n_{\sigma^\prime}}{n}\int
[g_{\sigma,\sigma^\prime}(r)-1]e^{-i\kk\cdot\rr}
\,d\rr+\frac{n_\sigma n_{\sigma^\prime}}{n}(2\pi)^3\delta(\kk)~,
\eq
From now on we will ignore the delta function at $\kk=0$. The total
(spin-summed) static structure factor is 
$S=\sum_{\sigma,\sigma^\prime}S_{\sigma,\sigma^\prime}$. Due to the
charge sum rule (\ref{0msr}) we must have $\lim_{k\to 0}S(k)=0$. In
Sec. \ref{sec:RPA} we will show that the small $k$ behavior of $S(k)$
has to start from the term of order $k^2$. 

\subsection{Analytic expressions for the non-interacting fermions}

Usually $g_{\sigma,\sigma^\prime}$ is conventionally divided into the
(known) exchange and the (unknown) correlation terms
\bq
g_{\sigma,\sigma^\prime}=g^x_{\sigma,\sigma^\prime}+g^c_{\sigma,\sigma^\prime}~,
\eq
where the exchange term corresponds to the uniform system of
non-interacting fermions.

\subsubsection{Radial distribution function}

We thus have (from the definition of the RDF
(\ref{RDF}) and using Slater determinants for the wave-function)
\bq \label{gx1}
g^x_{+,-}(r)&=&1~,\\ \label{gx2}
g^x_{\sigma,\sigma}(r)&=&1-\left[\frac{3j_1(k_F^\sigma r)}
{k_F^\sigma r}\right]^2~,
\eq
where $j_1(x)=[\sin(x)-x\cos(x)]/x^2$ is the spherical Bessel function
of the first kind and $(k^\sigma_F)^3=6\pi^2n_\sigma$ is the Fermi
wave-number for particles of spin $\sigma$.

\subsubsection{Static structure factor}

Again we will have the splitting
$S_{\sigma,\sigma^{\prime}}=S_{\sigma,\sigma^{\prime}}^x+S_{\sigma,\sigma^{\prime}}^c$
into the exchange and the correlation parts. So that for
the non-interacting fermions we get
\bq \label{Sx1}
S^x_{+,-}(k)&=&0~,\\ \nonumber
S^x_{\sigma,\sigma}(k)&=&\frac{n_\sigma}{n}-\frac{n_\sigma^2}{n}
\Theta(2k^\sigma_F-k)\frac{3\pi^2}{(k^\sigma_F)^3}
\left(1-\frac{k}{2k^\sigma_F}\right)^2\left(2+\frac{k}{2k^\sigma_F}\right)
\\ \label{Sx2}
&=&\frac{n_\sigma}{n} \left\{\begin{array}{ll}
1 & k>2k^\sigma_F\\
\frac{3}{4}\frac{k}{k^\sigma_F}-\frac{1}{16}
\left(\frac{k}{k^\sigma_F}\right)^3 & k<2k^\sigma_F
\end{array}
\right.~, 
\eq
where $\Theta(x)$ is the Heaviside step function. 

\subsection{RDF sum rules}

Both the behavior of the RDF at small $r$ and at large $r$ has to
satisfy to general exact relations or sum rules. 

\subsubsection{Cusp conditions}

When two electrons ($\mu=\infty$) get closer and closer together, the
behavior of $g_{\sigma,\sigma^\prime}(r)$ is governed by the exact
cusp conditions \cite{Kimball1973,*Rajagopal1978,*Hoffmann1992} 
\bq
\left.\frac{d}{dr}g_{\sigma,\sigma}(r)\right|_{r\to 0}&=&0~,\\
\left.\frac{d^3}{dr^3}g_{\sigma,\sigma}(r)\right|_{r\to 0}&=&
\frac{3}{2a_0}\left.\frac{d^2}{dr^2}g_{\sigma,\sigma}(r)\right|_{r\to 0}~,\\
\left.\frac{d}{dr}g_{+,-}(r)\right|_{r\to 0}&=&\frac{1}{a_0}
g_{+,-}(0)~,
\eq
where in the adimensional units $a_0\to 1/r_s$. For finite $\mu$ we
only have the condition $g_{\sigma,\sigma}(0)=0$ due to Pauli
exclusion principle.

\subsubsection{The Random Phase Approximation (RPA) and the long range
  behavior of the RDF}
\label{sec:RPA}

Within the linear density response theory \anew{\cite{Hansen,static}}
 one introduces
the space-time Fourier transform, $\chi(\kk,\omega)$, of the linear
density response function. Which is related through the fluctuation
dissipation theorem,
$S(\kk,\omega)=-(2\hbar/n)\Theta(\omega)\text{Im}\chi(\kk,\omega)$, to
the space-time Fourier transform, $S(\kk,\omega)$ (dynamic structure
factor), of the van Hove correlation function  \cite{vanHove54}, 
$\langle\rho(\rr,t)\rho(\mathbf{0},0)\rangle/n$, where
$\rho(\rr,t)=\exp(iHt/\hbar)\rho(\rr)\exp(-iHt/\hbar)$. 

In the Random Phase Approximation (RPA) we have \cite{Pines}
\bq
\frac{1}{\chi_{RPA}(k,\omega)}=
\frac{1}{\chi_0(k,\omega)}-e^2\tilde{v}_\mu(k)~,
\eq   
where $\chi_0$ is the response function of the non-interacting
Fermions (ideal Fermi gas), known
as the Lindhard susceptibility \cite{Lindhard54}. This corresponds to
taking the ``proper polarizability'' (the response to the Hartree
potential) equal to the response of the ideal Fermi gas
\cite{Tosi1999}. The RPA static structure factor is then recovered
from the fluctuation dissipation theorem as follows
 \bq
S_{RPA}(k)=-\frac{\hbar}{n}\int_0^\infty\frac{d\omega}{\pi}\,
\text{Im}\chi_{RPA}(k,\omega)~.
\eq
where
\bq
\text{Im}\chi_{RPA}=\frac{\text{Im}\chi_0}{(1-e^2\tilde{v}_\mu\text{Re}\chi_0)^2
+(e^2\tilde{v}_\mu\text{Im}\chi_0)^2}~,
\eq

The small $k$ behavior of the RPA structure factor is exact
\cite{Pines}. One finds  
\bq
S_{RPA}(k)=\frac{\hbar k^2}{2m\omega_p}~,~~~k\ll k_F~,
\eq
where $\omega_p=\sqrt{4\pi ne^2/m}$ is the plasmon frequency
\cite{Giuliani-Vignale}. This is also known as the second-moment sum
rule for the exact RDF and can be rewritten as $n\int
d\rr\,r^2[g(r)-1]=-6(\hbar/2m\omega_p)$. We can then say that $g(r)-1$
has to decay faster than $r^{-5}$ at large $r$. The fourth-moment (or
compressibility) sum rule links the thermodynamic compressibility,
\anew{$\chi=[nd(n^2de_0/dn)/dn]^{-1}$}, \cite{Tosi1999} to 
the fourth-moment of the RDF. For the equivalent
classical system it is well known that the correlation functions have
to decay faster than any inverse power of the distance
\cite{Alastuey1985,Martin88,Lighthill} 
(in accord with the Debye-H\"ukel theory). We are not aware of the
existence of a similar result for the zero temperature
quantum case. 

\section{Results of the calculation}
\label{sec:results}

We considered fourty systems corresponding to $r_s=1,2,5,10$,
$\mu=\infty,4,2,1,1/2$, $\zeta=0,1$. For each system we
calculated the RDF using the histogram estimator in a variational,
mixed, and extrapolated measure and a particular HFM measure. Before
starting with the simulations we determined the optimal values for the
time step $\tau$ and the number of walkers $n_w$ for each density.
 
\subsection{Extrapolations}
\label{sec:extra}
For the Coulomb case, $\mu\to\infty$, we made extrapolations in time
step $\tau$ and number of walkers $n_w$ for each value of
$r_s$ within our DMC simulations. Given a relative precision
$\delta_{e_0}=\Delta e_0/e_p^x $, where 
$e_0=\langle E_L\rangle_{f}/N$, $\Delta e_0$ is the statistical error on $e_0$,
and $e_p^x$ is the exchange energy per particle (see Eq. (\ref{epx})),
we set as our target relative precision $\delta_{e_0}=10^{-2}\%$.

\subsubsection{In time step}

Our results are summarized in Table \ref{tab2}.
As the characteristic dimension of one particle diffusing walk is
$\sigma=\sqrt{2\lambda\tau}$ or $\sqrt{2\tau/r_s^2}$ in adimensional
units, this has to remain 
of the order of the mean nearest neighbor separation $a$ which is
chosen to be a constant in our units. Then we expect that at lower
$r_s$ one needs to choose smaller time steps $\tau$. For this reason
we chose different time steps in the simulations of the Table:
$\tau=0.5,0.1,0.05$ for $r_s=10$, $\tau=0.3,0.1,0.05$ for $r_s=5$,
$\tau=0.05,0.03,0.005$ for $r_s=2$, and $\tau=0.01,0.005,0.001$ for
$r_s=1$. Note that, at fixed $r_s$, the statistical errors increase as
the time step diminishes. 

\begin{table}
\caption{Extrapolation in time step for $N=66$ unpolarized
  electrons $(\mu=\infty)$ at a fixed number of $n_w=600$ walkers
  with a trial wave-function of the SJ type. We run the
  simulation for 3 different time steps and did 
  a linear fit of the $(\tau,e_0)$ data, $e_0=a+b\tau$. The optimal
  $\tau$  is the largest one compatible 
  with the target precision. \label{tab2}}  
{\scriptsize
\begin{ruledtabular}
\begin{tabular}{ccccc}
$r_s$&$a$&$b$&$\chi^2$&optimal $\tau$\\
\hline
10&-0.107456(7)&0.00010(2)&0.9&0.09\\
5&-0.153352(4)&0.00024(3)&0.1&0.07\\
2&-0.00416(8)&0.003(2)&4.4&0.01\\
1&1.14579(7)&0.032(9)&1.1&0.003\\
\end{tabular}
\end{ruledtabular}
}
\end{table}

\subsubsection{In the number of walkers}

Our results are summarized in Table \ref{tab3}. The fluctuations of
the statistical weight of a walker depend on the fluctuations of
the local energy, {\sl i.e.} by the quality of the trial
wave-function. The quality of the trial wave-function worsens as $r_s$
becomes larger (for the strongly correlated system), and one expects
that the necessary number of walkers increases. This is in agreement
with the results of the Table. Note that, at fixed $r_s$, the
statistical errors increase as the number of walkers diminishes.  

\begin{table}
\caption{Extrapolation in number of walkers for $N=66$ unpolarized
  electrons $(\mu=\infty)$ with a time step $\tau=0.1$ for
  $r_s=10,5$, $\tau=0.05$ for $r_s=2$, and $\tau=0.01$ for $r_s=1$
  with a trial wave-function of the SJ type. We run the 
  simulation for 4 different numbers of walkers and did 
  a linear fit of the $(1/n_w,e_0)$ data, $e_0=a+b/n_w$. The optimal $n_w$
  is the smallest one compatible 
  with the target precision. \label{tab3}} 
{\scriptsize
\begin{ruledtabular}
\begin{tabular}{ccccc}
$r_s$&$a$&$b$&$\chi^2$&optimal $n_w$\\
\hline
10&-0.107443(3)&0.0032(4)&0.1&354\\
5&-0.153329(6)&0.0044(7)&0.2&243\\
2&-0.004036(6)&0.0026(7)&0.2&56\\
1&1.14609(6)&0.01(1)&1.2&40\\
\end{tabular}
\end{ruledtabular}
}
\end{table}

\subsection{Effect of backflow and three-body correlations}

\begin{figure}[H]
\begin{center}
\includegraphics[width=8cm]{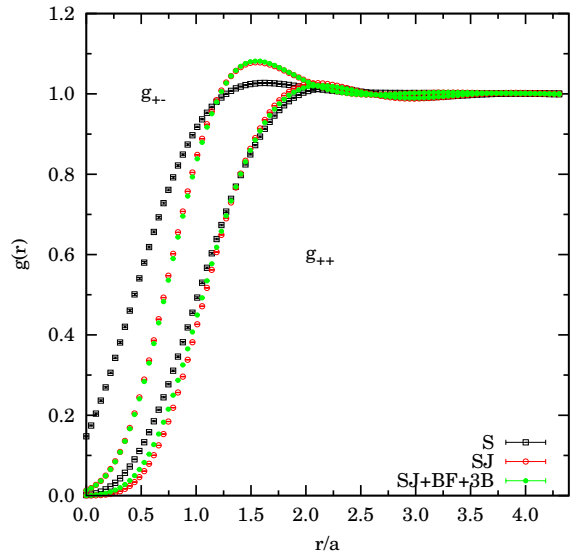}
\end{center}
\caption{Shows the mixed measure of the RDF calculated in DMC for
  $N=162, \zeta=0, \mu=\infty, r_s=10$ with a S, SJ, SJ+BF+3B trial
  wave-function.} 
\label{fig:grwf}
\end{figure}

In Fig. \ref{fig:grwf} we show the mixed measure of the RDF
calculated in DMC for $N=162, 
\zeta=0, \mu=\infty, r_s=10$ with different kinds of trial
wave-functions. Of course in a VMC calculation using the Slater 
determinant wave-function gives us $g^x_{\sigma,\sigma^\prime}$, the
RDF of the ideal gas (see Eqs. (\ref{gx1})-(\ref{gx2})).

\begin{figure}[H]
\begin{center}
\includegraphics[width=8cm]{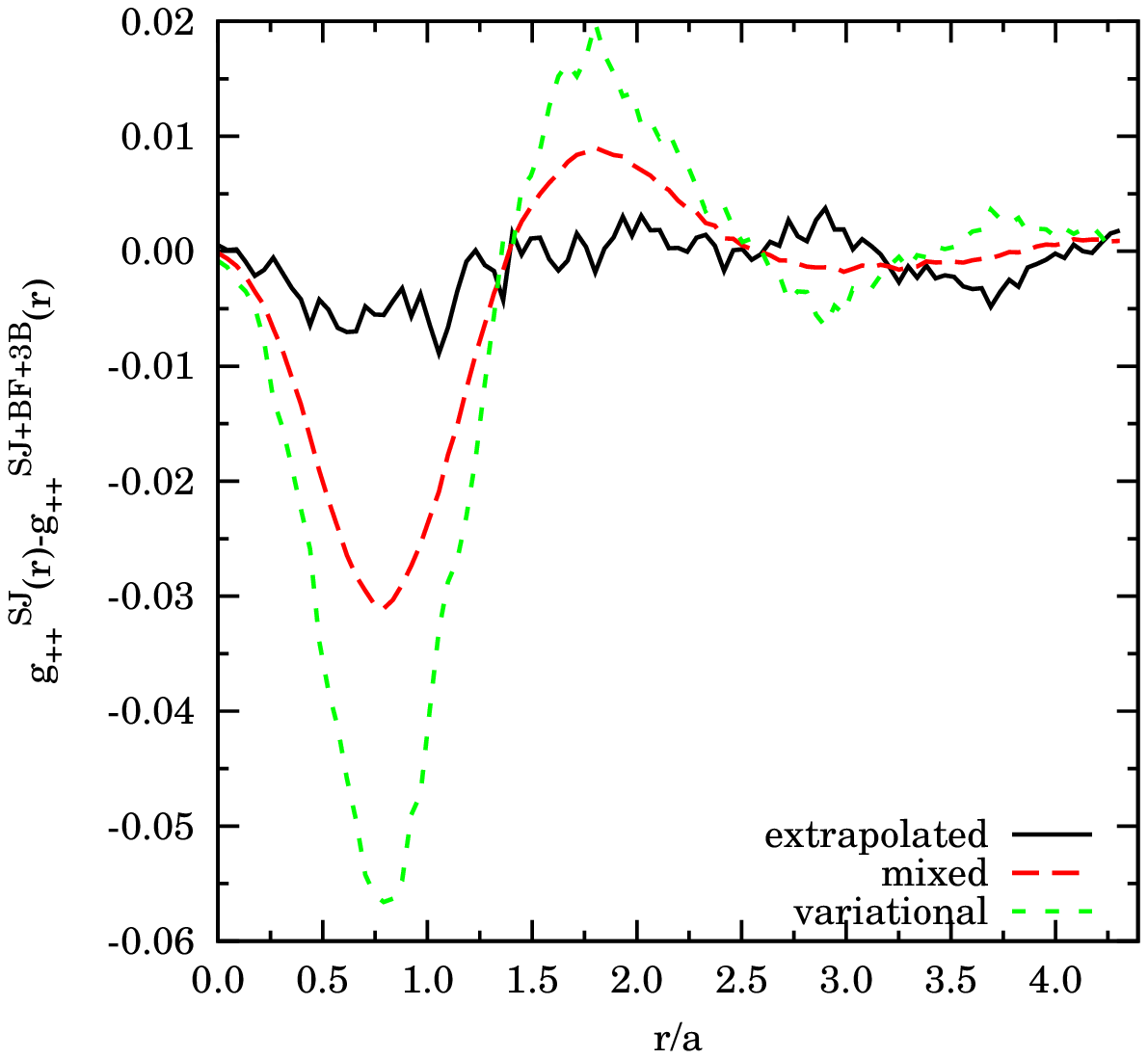}
\includegraphics[width=8cm]{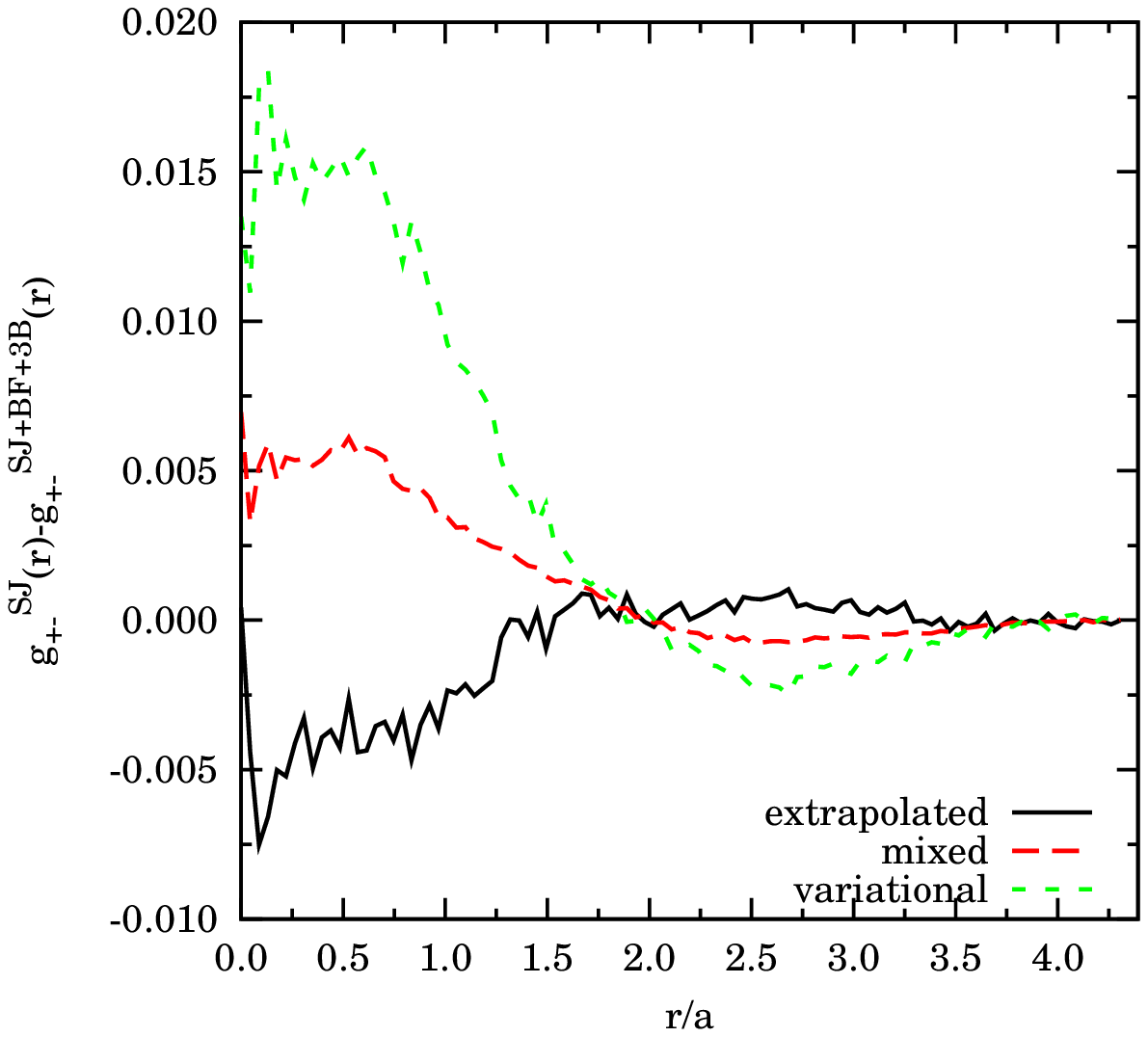}
\end{center}
\caption{Shows the difference between the RDF calculated with the SJ
  wave-function and the one calculated with the SJ+BF+3B wave-function
  using the variational, the mixed, and the extrapolated measure. The
  results are for 
  $N=162, \zeta=0, \mu=\infty$. On the left the like RDF is used at
  $r_s=10$, on the right the unlike RDF at $r_s=1$.}
  \label{fig:dgr}
\end{figure}

In Fig. \ref{fig:dgr} we show the difference between the RDF
calculated with the SJ wave-function and the one calculated with the
SJ+BF+3B wave-function using the variational, the mixed, and the
extrapolated measure. 

With the extrapolated measure the results from the SJ computation
differs by less than $0.005$ from the ones from the SJ+BF+3B. We then
decided to perform our subsequent calculations using the SJ trial
wave-function. 

\subsection{Size effects}

\begin{figure}[H]
\begin{center}
\includegraphics[width=8cm]{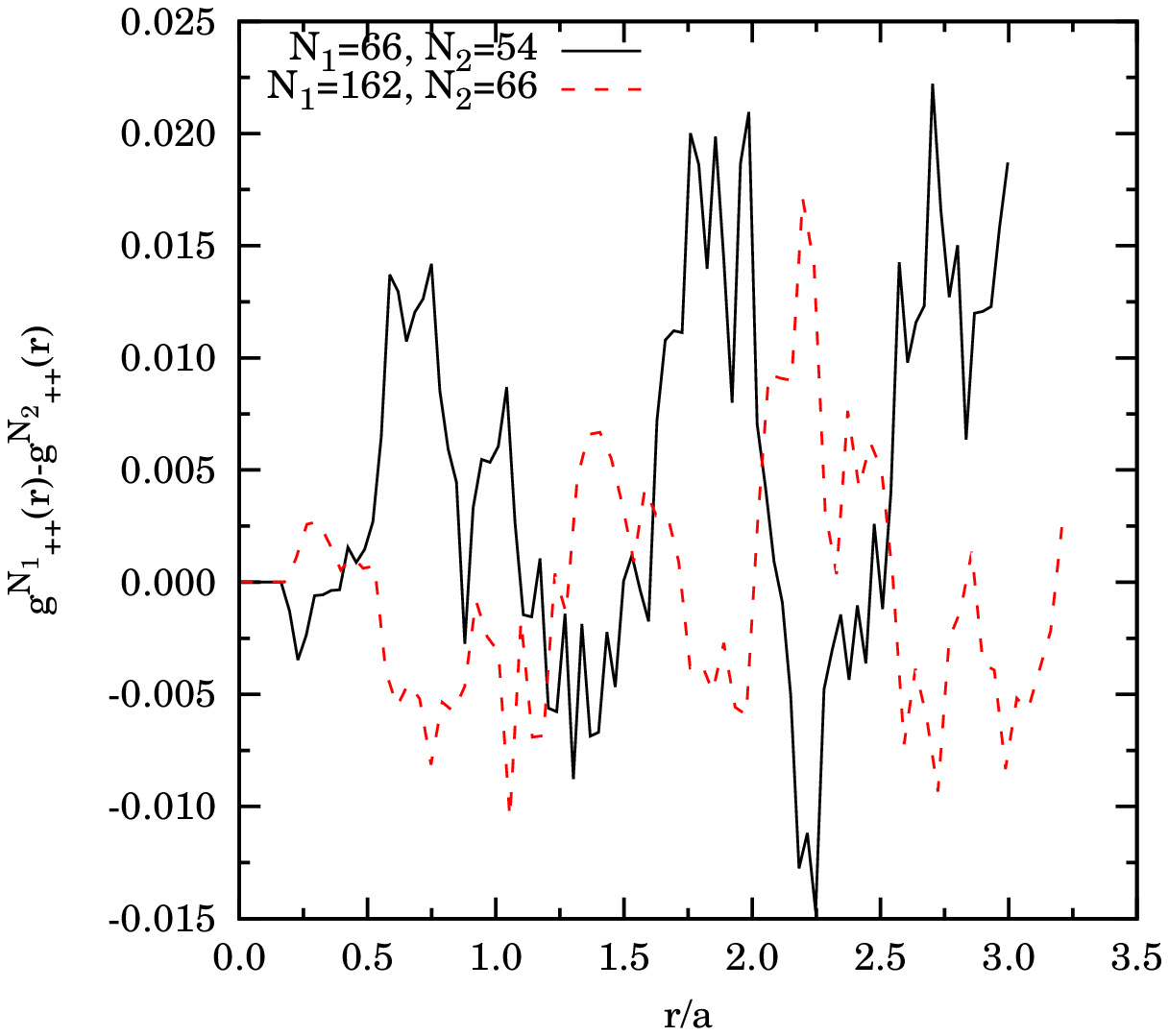}
\includegraphics[width=8cm]{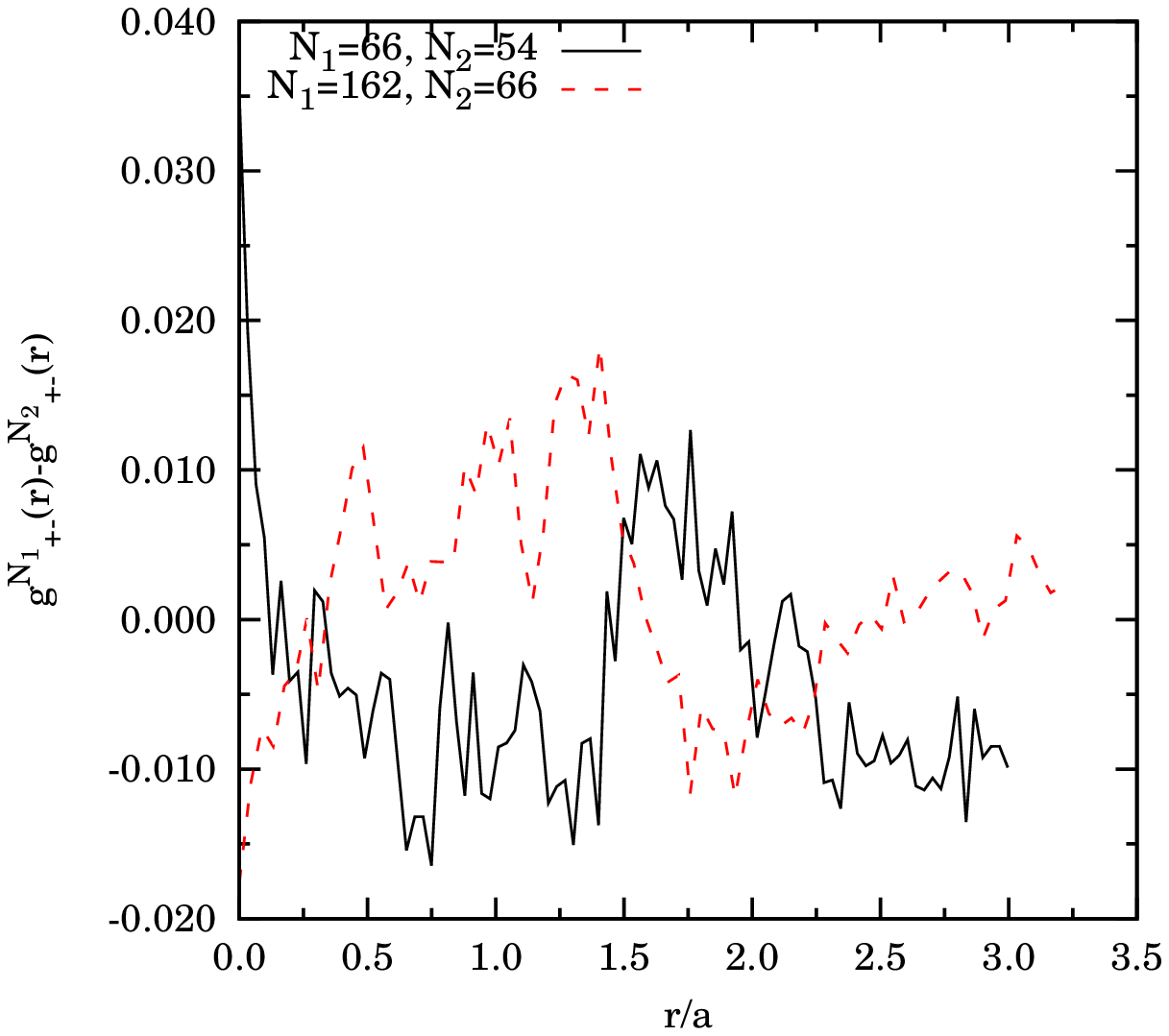}
\end{center}
\caption{Shows the difference between the RDF of two systems of
  electrons ($\mu=\infty$) at $r_s=10$ and $\zeta=0$ with different
  sizes $N_1$ and $N_2$. The RDF are calculated in VMC with
  the SJ wave-function. On the left the difference of the like RDF is
  shown and on the right the difference of the unlike RDF is shown.}
\label{fig:grn}
\end{figure}

In order to estimate the size effects on the RDF calculation we
performed a series of VMC calculation with the SJ
wave-function on an unpolarized system with 
different number of particles. The results (see Fig. \ref{fig:grn})
show that the size dependence mainly affects the long range behavior
of the RDF and the on-top value for the unlike one. 
In the simulation the RDF is defined on $r\in[0,r_{max}]$
with $r_{max}=L/2$ where $L=\Omega^{1/3}=(4\pi N/3)^{1/3}$ is the size
of the simulation box. To minimize size effects we chose to
perform our RDF calculation with $N=162$ in the 
unpolarized case and $N=147$ in the polarized case.

\subsection{The HFM measure}
\label{sec:ZVZBr}

From the definition (\ref{RDFhi}), we can write the RDF as 
\bq
g_{\sigma,\sigma^\prime}(r)=\frac{\langle
  I_{\sigma,\sigma^\prime}(r,\RR)\rangle}{\Omega n_\sigma n_{\sigma^\prime}}~.
\eq
Since the operator $I_{\sigma,\sigma^\prime}$ is diagonal in
coordinate representation then
$I_{\sigma,\sigma^\prime}=(I_{\sigma,\sigma^\prime})_L$.  
Indicating with $\Omega_\rr$ the solid angle spanned by the
$\rr$ vector, we can write
\bq
I_{\sigma,\sigma^\prime}(r,\RR)=\sum_{i,j\neq i}\delta_{\sigma,\sigma_i}
\delta_{\sigma^\prime,\sigma_j}\int\frac{d\Omega_\rr}{4\pi}
\delta(\rr-\rr_{ij})~,
\eq
which is the usual histogram estimator \cite{Allen-Tildesley}.
Following Toulouse \cite{Toulouse2007} we choose for $Q$ the following
expression  
\bq \label{QT}
Q_{\sigma,\sigma^\prime}(r,\RR)=
-\frac{r_s^2}{8\pi}\sum_{i,j\neq i}\delta_{\sigma,\sigma_i}
\delta_{\sigma^\prime,\sigma_j}\int\frac{d\Omega_\rr}{4\pi}
\frac{1}{|\rr-\rr_{ij}|}~,
\eq
so that (using the identities $\nablab^2_{\rr_{ij}}1/|\rr-\rr_{ij}| =
-4\pi\delta(\rr-\rr_{ij})$ and $\nablab_{\rr_i}f(\rr_{kj}) =
\nablab_{\rr_{kj}}f(\rr_{kj})[\delta_{ik}-\delta_{ij}]$, for a given
function $f$) the first
term in Eq. (\ref{zvq}) exactly cancels the histogram 
estimator $I_{\sigma,\sigma^\prime}$. Then the HFMv estimator is
\bq \nonumber
I_{\sigma,\sigma^\prime}^\text{HFMv}(r,\RR)&=&
\frac{1}{2\pi}\sum_{i,j\neq i}\delta_{\sigma,\sigma_i}
\delta_{\sigma^\prime,\sigma_j}\vv_i(\RR)\cdot\int\frac{d\Omega_\rr}{4\pi}
\nablab_{\rr_{ij}}\frac{1}{|\rr-\rr_{ij}|}\\ \label{ZVgr}
&=&-\frac{1}{4\pi}
\sum_{i,j\neq i}\delta_{\sigma,\sigma_i}\delta_{\sigma^\prime,\sigma_j}
\vv_i(\RR)\cdot\frac{\rr_{ij}}{r_{ij}^3}[1+\sgn(r_{ij}-r)]~,
\eq 
which goes to zero at large $r$ (see Note \onlinecite{shift}).
The correct (taking care of the missing factor of two in Ref.
\onlinecite{Toulouse2007}) $\beta$ correction is 
\bq\nonumber
\Delta I_{\sigma,\sigma^\prime}^{\beta}(r,\RR)&=&-[E_L(\RR)-E_0]\frac{r_s^2}{8\pi}
\sum_{i,j\neq i}\delta_{\sigma,\sigma_i}
\delta_{\sigma^\prime,\sigma_j}\int\frac{d\Omega_\rr}{4\pi}
\frac{1}{|\rr-\rr_{ij}|}\\ \label{ZBgr}
&=&-[E_L(\RR)-E_0]\frac{r_s^2}{16\pi}
\sum_{i,j\neq i}\delta_{\sigma,\sigma_i}
\delta_{\sigma^\prime,\sigma_j}
\left(\frac{r_{ij}+r-|r_{ij}-r|}{r_{ij}r}\right)~.
\eq
Note that also $\langle\Delta
I^{\beta}_{\sigma,\sigma^\prime}(r,\RR)\rangle$ goes to zero at large
$r$. This particular HFM measure needs to be shifted
$g_{\sigma,\sigma^\prime}(r) = 
g^\text{HFM}_{\sigma,\sigma^\prime}(r)+1$. We chose to do the shift as 
follows: $g_{\sigma,\sigma^\prime}(r)=g^\text{HFM}_{\sigma,\sigma^\prime}(r)
+g^\text{mix}_{\sigma,\sigma^\prime}(L/2)-g^\text{HFM}_{\sigma,\sigma^\prime}(L/2)$. 
Nonetheless it is expected to give better results for the on-top value
of the RDF where the histogram estimator of Eq. (\ref{RDFhi}), after
the necessary discretization of the Dirac delta function, leads, in
the measure, to a statistical average divided by zero. Moreover it
does not suffer from any discretization error and can be calculated
for any value of $r$.

\begin{figure}[H]
\begin{center}
\includegraphics[width=8cm]{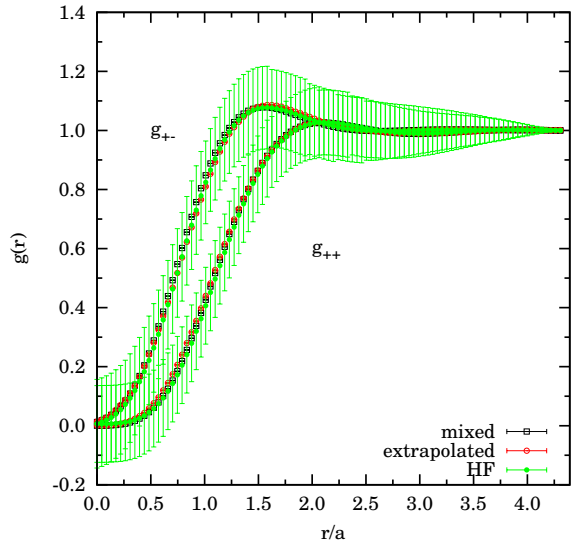}
\end{center}
\caption{Shows the RDF of the $N=162,
\zeta=0, \mu=\infty, r_s=10$ system, calculated in DMC SJ with various
kinds of measures: mixed histogram (mixed), extrapolated histogram 
(extrapolated), and HFM (HFM) with the choice of Eq. (\ref{QT}).} 
\label{fig:gre}
\end{figure}

In Fig. \ref{fig:gre} we show a comparison for the RDF of the $N=162,
\zeta=0, \mu=\infty, r_s=10$ system, calculated in DMC SJ with various
kinds of measures. The length of the run was always the same 50
blocks of 500 steps each. From the figure one can see that with our
choice of the $\beta$ correction the HFM measure has the correct
average value (coinciding with the usual histogram estimator). From
the figure it is also evident that the HFM measure is much less
efficient than the other measures (clearly with a sufficient number of
blocks the statistical error on the HFM measure can be made small
at will). 
This inefficiency is entirely due to the ZB correction (essential in the DMC
calculation). From its definition (see Eq. (\ref{ZBgr})) one can see
that it is the small difference of two large terms involving the
(extensive) total energy . So the statistical error on the HFM
measure is completely dominated by that of the $\beta$ part, the
$\alpha$ part having statistical errors comparable with the ones of
the usual histogram estimator, as shown in the left panel of
Fig. \ref{fig:cusp}.

\subsection{Choice of the Jastrow}
\label{sec:jastrow-results}

\begin{figure}[H]
\begin{center}
\includegraphics[width=8cm]{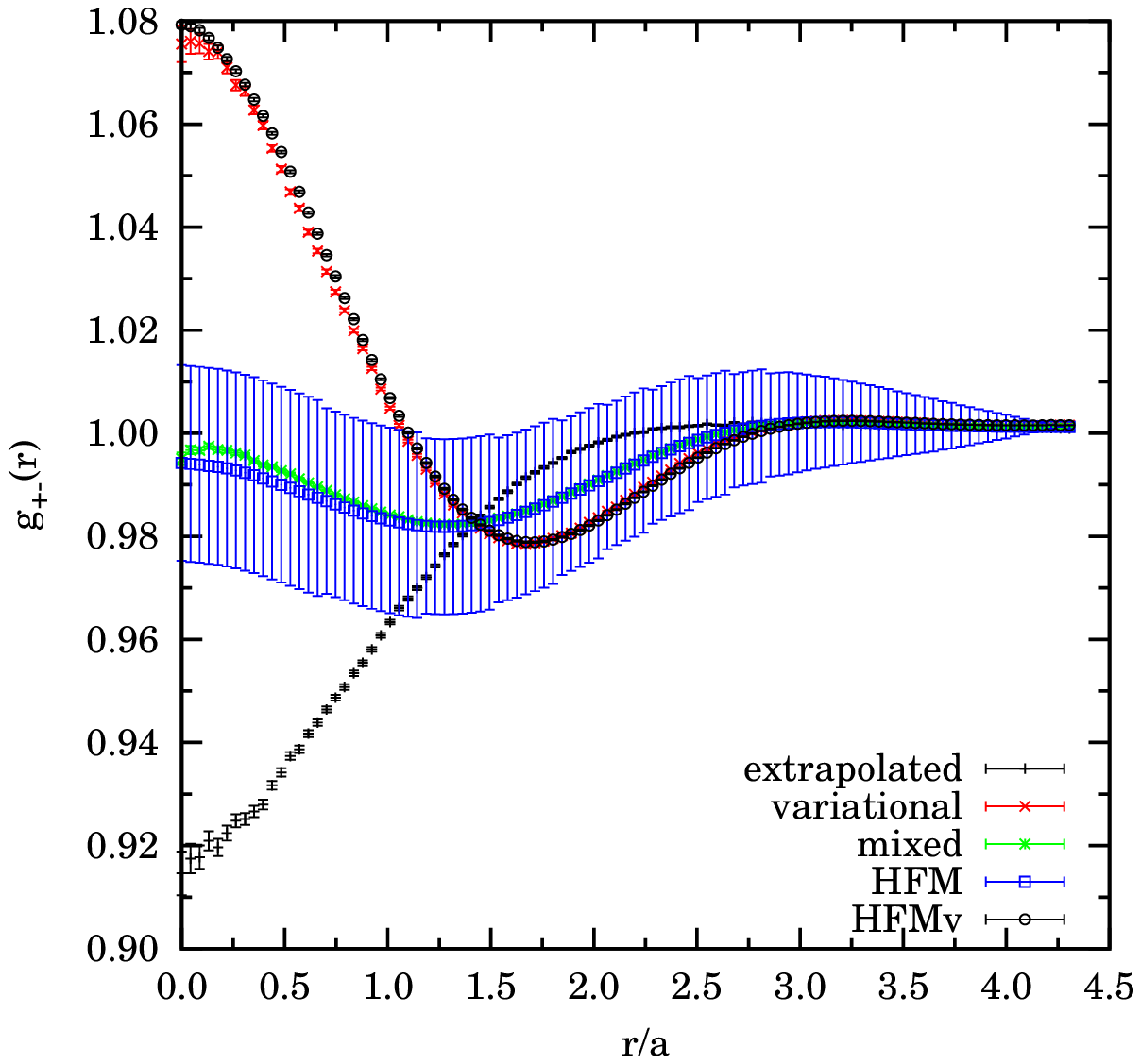}
\includegraphics[width=8cm]{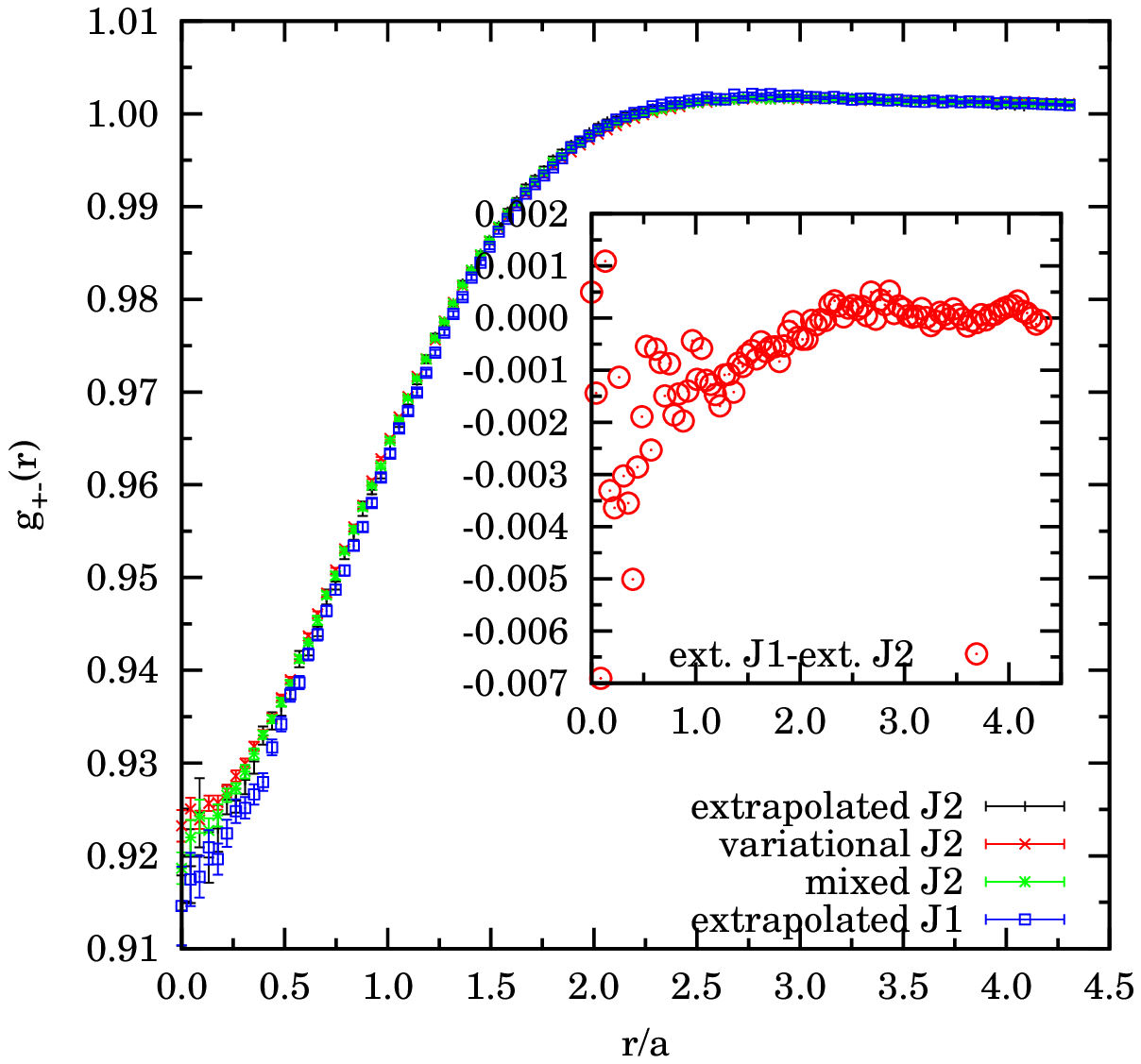}
\end{center}
\caption{Unlike RDF for the unpolarized fluid of Paziani
  \cite{Paziani2006} at $r_s=1$ and $\mu=1$ with
  $N=162$. On the left the calculation with the Jastrow ${\cal J}_1$ of
  Eq. (\ref{jastrowrpam}) with various measures: variational
  histogram (variational) and variational HFMv (HFMv) using the estimator
  of Eq. (\ref{ZVgr}), mixed histogram (mixed) and HFM (HFM), and
  extrapolated histogram (extrapolated). On the right the
  calculation with the Jastrow ${\cal J}_2$ of Eq. (\ref{jastrowrpa}) with the
  histogram variational (variational), mixed (mixed), and extrapolated
  (extrapolated) measures. In the inset is
  shown the difference between the histogram extrapolated measure
  of the calculation with ${\cal J}_1$ and the histogram extrapolated
  measure of the calculation with ${\cal J}_2$. $10^5$ Monte Carlo steps were
  used in the simulations.}
\label{fig:cusp}
\end{figure}

We noticed that at small $r_s, \mu,$ and $r$ the variational measure
for the unlike RDF, with the chosen Jastrow ${\cal J}_1$ of
Eq. (\ref{jastrowrpam}), deviates strongly from the mixed 
one. This is no longer so with the modified Jastrow
${\cal J}_2$ of Eq. (\ref{jastrowrpa}) which at small $\mu$ gives also better
variational energies (but not for $\mu\to\infty$ where ${\cal J}_1$ is
better. Note that the 
Jastrow factor does not change the nodes of the wave-function so the
energies calculated from the diffusion with ${\cal J}_1$ or ${\cal
  J}_2$ coincide). The 
extrapolated measures do not change appreciably in the two cases
apart from near $r=0$. In Fig. \ref{fig:cusp} we show the difference
for the two calculations with ${\cal J}_1$ and ${\cal J}_2$ for the
$\zeta=0, r_s=1, \mu=1$ 
model. From the inset in the left panel we can see that among the two
extrapolated measures there is a difference of the order of 0.005.   

Our results with the two Jastrow factors show that ${\cal J}_1$ is better than
${\cal J}_2$ for the near-Jellium systems ($\mu$ large) while ${\cal
  J}_2$ is better than ${\cal J}_1$ for the near-ideal systems ($\mu$
small).  

\subsection{The histogram estimator}

\begin{figure}[H]
\begin{center}
\includegraphics[width=14cm]{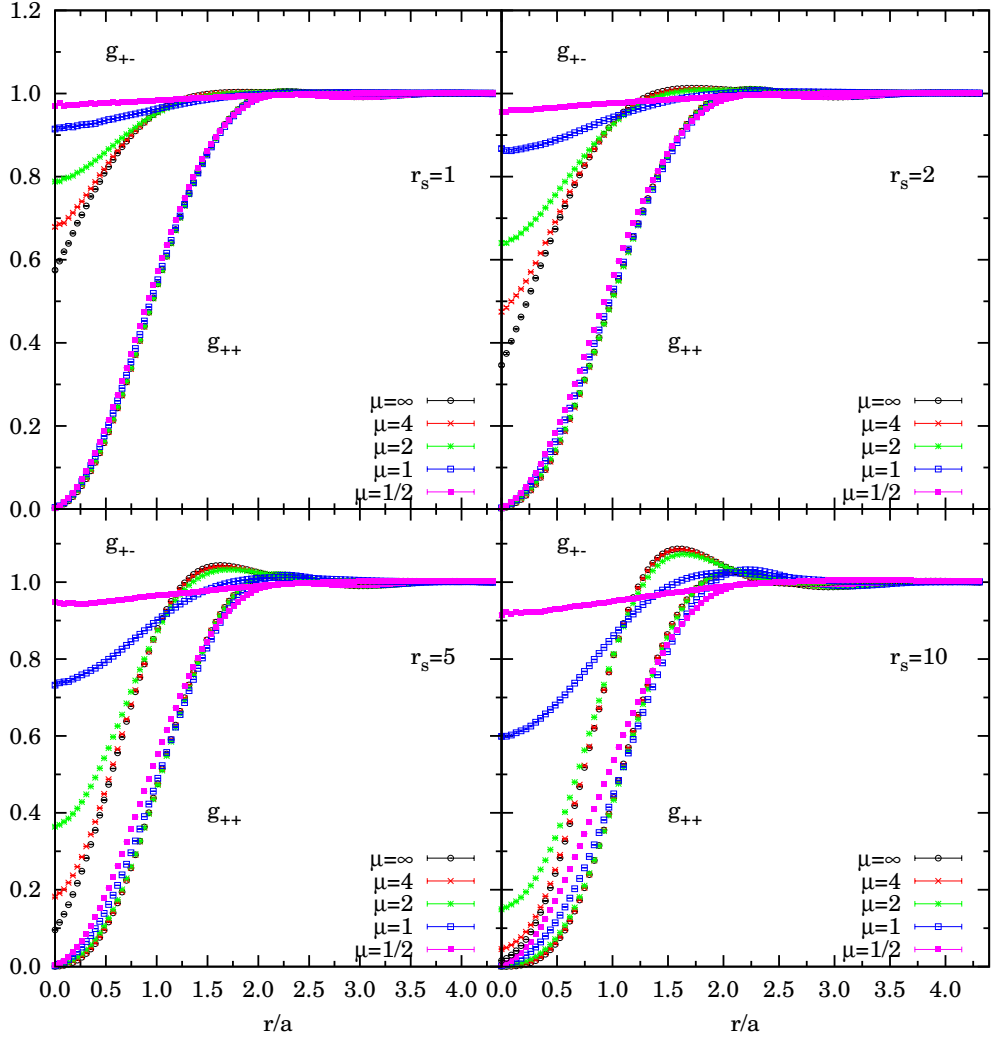}
\end{center}
\caption{Shows the histogram extrapolated measure for the RDF of a
  system of 
  162 unpolarized ($\zeta=0$) particles calculated using the SJ trial
  wave-function. The VMC calculation was made of $10^6$ steps while 
  the DMC by $10^5$. The trial wave-function used was of the
  SJ type with the Jastrow ${\cal J}_1$ of Eq. (\ref{jastrowrpam}).}
\label{fig:gr_z0}
\end{figure}

\begin{figure}[H]
\begin{center}
\includegraphics[width=14cm]{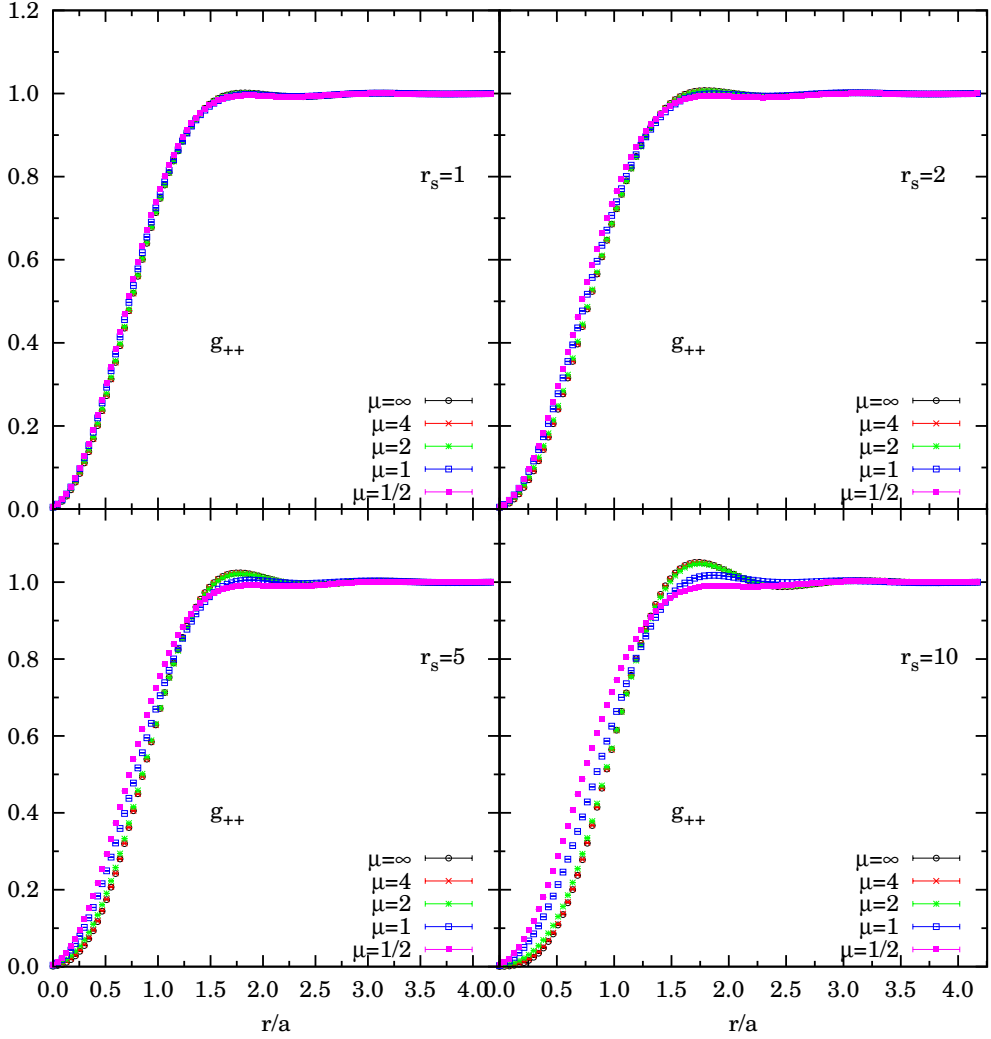}
\end{center}
\caption{Shows the histogram extrapolated measure for the RDF of a
  system of 
  147 fully polarized ($\zeta=1$) particles calculated using the SJ trial
  wave-function. The VMC calculation was made of $10^6$ steps while
  the DMC by $10^5$. The trial wave-function used was of the
  SJ type with the Jastrow ${\cal J}_1$ of Eq. (\ref{jastrowrpam}).}
\label{fig:gr_z1}
\end{figure}

In Fig. \ref{fig:gr_z0} we show the DMC results for the histogram
extrapolated measure of the RDF of our
fluid model at $\zeta=0$. The time step, $\tau$, and number of
walkers, $n_w$, were chosen according to the indications given in
subsection \ref{sec:extra}. Fig. \ref{fig:gr_z1} is for the $\zeta=1$
case. 


In Table \ref{tab:got_z0} we show the on-top values for the unlike
RDF, $g_{+-}(0)$, of the unpolarized system, calculated with the
histogram variational, the histogram mixed, the histogram
extrapolated measure, the HFM measure, and the HFM extrapolated measure (of
Eq. (\ref{zvzb-e})). 

\begin{table}
\caption{Contact values for the unlike RDF of the unpolarized fluid of
  Paziani \cite{Paziani2006},
  at various $r_s$ and $\mu$, from the histogram variational
  (variational), mixed (mixed), and 
  extrapolated (extrapolated) measures, and the HFM (HFM) and HFM
  extrapolated (HFM-ext) measures. The trial 
  wave-function used was of the SJ type with the Jastrow ${\cal J}_1$ of
  Eq. (\ref{jastrowrpam}). The last column gives the
  error $\sigma_\text{av}=\sqrt{\sigma^2{\cal K}/{\cal N}}$ ($\sigma^2$
  is the variance, ${\cal K}$ the correlation time of the random walk,
  and ${\cal N}$ the number of Monte Carlo steps) on the HFM measure. 
  $162$ particles were used with $10^5\times n_w$ Monte Carlo steps.  
\label{tab:got_z0}}
{\scriptsize
\begin{ruledtabular}
\begin{tabular} {cccccccc}
$r_s$&$\mu$& variational & mixed & extrapolated & HFM & HFM-ext &
  $\sigma_{av}$ on HFM\\ 
\hline
10&1/2&1.085(8)&1.000(4)&0.91(1)&1.0006&0.9222&0.03\\
10&1&0.706(6)&0.644(3)&0.582(8)&0.6474&0.5949&0.03\\
10&2&0.219(4)&0.182(1)&0.146(4)&0.1798&0.1450&0.06\\
10&4&0.053(2)&0.0506(8)&0.048(2)&0.0460&0.0394&0.07\\
10&$\infty$&0.0074(6)&0.0096(3)&0.0118(8)&0.0045&0.0029&0.09\\
\hline
5&1/2&1.129(8)&1.034(3)&0.94(1)&1.0277&0.9381&0.03\\
5&1&0.850(7)&0.796(3)&0.743(9)&0.7912&0.7325&0.02\\
5&2&0.448(5)&0.405(2)&0.362(6)&0.4022&0.3565&0.02\\
5&4&0.214(3)&0.199(1)&0.184(4)&0.1960&0.1782&0.03\\
5&$\infty$&0.080(2)&0.0799(8)&0.080(2)&0.0625&0.0557&0.03\\
\hline
2&1/2&1.158(8)&1.0618(4)&0.97(1)&1.0545&0.9484&0.04\\
2&1&1.003(8)&0.927(3)&0.852(9)&0.9270&0.8561&0.03\\
2&2&0.754(7)&0.697(3)&0.639(9)&0.6919&0.6299&0.02\\
2&4&0.549(6)&0.511(2)&0.473(7)&0.5127&0.4687&0.02\\
2&$\infty$&0.376(4)&0.349(2)&0.323(5)&0.3236&0.3030&0.02\\
\hline
1&1/2&1.171(8)&1.077(3)&0.98(1)&1.0705&0.9683&0.02\\
1&1&1.077(8)&0.994(3)&0.91(1)&0.9938&0.9070&0.02\\
1&2&0.924(8)&0.855(3)&0.787(9)&0.8640&0.8053&0.02\\
1&4&0.784(7)&0.730(2)&0.676(8)&0.7295&0.6628&0.01\\
1&$\infty$&0.645(6)&0.602(2)&0.560(7)&0.5771&0.5263&0.01\\
\end{tabular}
\end{ruledtabular}
}
\end{table}
%

\section{Conclusions}
\label{sec:conclusions}

We studied through Variational and Diffusion Monte Carlo techniques
the fluid of spin one-half particles interacting with the bare pair-potential
$v_\mu(r)=\text{erf}(\mu r)/r$ and immersed in a uniform counteracting
background. When $\mu\to\infty$ the system reduces to the Jellium
model whereas when $\mu\to 0$ it reduces to the ideal Fermi gas. We
performed a detailed analysis of the spin-resolved Radial Distribution
Function for this system as a function of the density parameter
$r_s=1,2,5,10$ and the penetrability parameter $\mu=1/2,1,2,4,\infty$
at two values of the polarization, $\zeta=0,1$.  

Initially we carefully fine tuned our DMC calculation determining the
optimal values for the time step $\tau$ and the number of walkers
$n_w$ for each value of the density parameter $r_s$. Increasing the
system size $N$ the RDF extends its range $[0,r_{max}]$ at larger
$r_{max}$. We estimated that for $N\ge 66$ the size dependence of the
RDF is lower than 2\%. As a compromise between computational cost and
reduction of the size effects, the largest uncontrolled source of
uncertainty on our RDF measurements, we chose to perform the RDF
calculation with $N=162$ in the unpolarized case and $N=147$ in the
polarized case. 

We calculated the RDF using two different routes: through the usual
histogram estimator and through a particular HFM measure. As expected,
in the VMC calculations the HFMv estimator gives better results for the
on-top value of the RDF. In the DMC calculation the inclusion of the
$\beta$ correction (which must be omitted in the VMC calculation) is
indispensable. Moreover the ZV estimator is zero for $r>r_{max}$ so
it has to be shifted by $+1$. 
From our variational and fixed nodes diffusion Monte Carlo experiments
turns out that although in the variational measure the average of the
histogram estimator agrees with the average of the HFMv estimator
within the square root of the variance of the average
$\sigma_\text{av}=\sqrt{\sigma^2{\cal K}/{\cal N}}$, where $\sigma^2$
is the variance, ${\cal K}$ the correlation time of the random walk,
and ${\cal N}$ the number of Monte Carlo steps, and the two
$\sigma_\text{av}$ are comparable, in 
the diffusion experiment, where one has to add the $\beta$ correction
not to bias the average, the Hellmann and Feynman measure has an
average in agreement with the one of the histogram estimator but the
$\sigma_\text{av}$ increases. This is to be expected from the
extensive nature of the 
$\beta$ correction in which the energy appears. Of course the averages
from the extrapolated Hellmann and Feynman measure and the extrapolated
measure for the histogram estimator also agree. 

In the simulation, for the Coulomb case, $\mu\to\infty$, we made
extrapolations in time step and number of walkers for each value of
$r_s$. Given a relative precision $\delta_{e_0}=\Delta e_0/e_p^x $, where 
$e_0=\langle E_L\rangle_{f}/N$, $\Delta e_0$ is the statistical error on $e_0$,
and $e_p^x$ is the exchange energy, we set as our target relative
precision $\delta_{e_0}=10^{-2}\%$. The extrapolated values of the
time step and number of walkers was then used for all other values of
$\mu$. We chose the trial wave function of the Bijl-Dingle-Jastrow
\cite{Bijl1940} form as a product of Slater determinats and a Jastrow
factor. The pseudo potential was chosen as in
Ref. \cite{Ceperley2004}, ${\cal J}_2$, 
which is expected to give better results for Jellium. Comparison with
the simulation of the unpolarized fluid at $r_s=1$ and $\mu=1$ with
the pseudo potential of Ref. \cite{Ceperley78}, ${\cal J}_1$, for
which the trial wave 
function becomes the exact ground state wave function in the $\mu\to
0$ limit, show that the two extrapolated measures of the unlike
histogram estimator differ one from the other by less than $7\times 10^{-3}$,
the largest difference being at contact (see Fig. 1). The use of more
sophisticated trial wave functions, taking into account the effect 
of backflow and three-body correlations, is found to affect the
measure by even less in the range of densities considered. For the
same reason we discarded the use of the twist-averaged boundary
conditions \cite{Lin2001} and only worked with periodic boundary
conditions. In Table \ref{tab:got_z0} we compare the contact 
values of the unlike RDF of the unpolarized fluid at various $r_s$ and
$\mu$ from the measures of the histogram estimator and the HFM
measures. We see that there is disagreement between the measure from
the histogram estimator and the HFM measure only in the Coulomb
$\mu\to\infty$ case at $r_s=1,2$.

Our results complement the ones of Paziani {\sl et al.}
\cite{Paziani2006} which only reported a limited number of RDF
data. We plan, in the future, to complete the calculation at
intermediate polarizations, $0<\zeta<1$, complementing the
work of Ortiz and Ballone \cite{Ortiz1994}, and Ceperley and coworkers
\cite{Kwon1998}.
 
We believe it is still an open problem the one of determining the
relationship between the choice of the auxiliary function, the bias it
introduces in the Hellmann and Feynman measure, and the variance of
this measure. 

\appendix
\section{Jastrow, backflow, and three-body} 
\label{app:bf+3bd}

In terms of the stochastic process governed by $f(\RR,t)$ one can
write, using Kac theorem \cite{Kac59,Kac51}
\bq \label{fnorm}
\int d\RR\, f(\RR,\tau)=\left\langle\exp\left[-\int_0^\tau
  dt\,E_L(\RR^t)\right]\right\rangle_{\text{DRW}}~,
\eq
where $\langle\ldots\rangle_{\text{DRW}}$ means averaging with respect
to the diffusing and drifting random walk. 
Choosing a complete set of orthonormal wave-functions $\Psi_i$
we can write for the true time dependent many-body wave-function 
\bq \nonumber
\phi(\RR,\tau)&=&\sum_i\Psi_i(\RR)\int
d\RR^\prime\Psi_i(\RR^\prime)\phi(\RR^\prime,\tau)\approx
\Psi(\RR)\int d\RR\, f(\RR,\tau)\\
&=& \Psi(\RR) \left\langle\exp\left[-\int_0^\tau
  dt\,E_L(\RR^t)\right]\right\rangle_{\text{DRW}}~,
\eq
where $\Psi$ is the wave-function, of the set, of maximum overlap with
the true ground-state, the trial wave-function. Assuming that at time
zero we are already close to the stationary solution, for sufficiently
small $\tau$ we can approximate 
\bq
\left\langle\exp\left[-\int_0^\tau
  dt\,E_L(\RR^t)\right]\right\rangle_{\text{DRW}}\approx
e^{-\tau E_L(\RR^\tau)}~.
\eq
By antisymmetrising we get the Fermion wave-function
\bq \label{lem}
\phi_F(\RR,\tau)\approx {\cal A}\left[e^{-\tau E_L(\RR)}
\Psi(\RR)\right]~,
\eq
where given a function $f(\RR)$ we define the operator (a symmetry of
the Hamiltonian)
\bq
{\cal A}[f(\RR)]=\frac{1}{N_P}\sum_P(-1)^Pf(P\RR)~,
\eq
here $N_P=N_+!N_-!$ is the total number of allowed permutations $P$.

This is called the local energy method to improve a trial
wave-function. Suppose we start from a simple unsymmetrical product of 
single particle plane waves of $N_+$ spin-up particles with $k<k_F^+$
occupied and  $N_-$ spin-up particles with $k<k_F^-$ occupied, for the
zeroth order trial wave-function. Equation (\ref{lem}) will give us
a first order wave-function of the Slater-Jastrow type (see equation
(\ref{twf})). If we 
start from an unsymmetrical Hartree-Jastrow trial wave-function the
local energy with the Jastrow factor has the form  
\bq \label{jel}
E_L=V-\lambda\sum_i\left[-k_i^2-2i\kk_i\cdot\nablab_i
\sum_{j<k}u(r_{jk})-\nablab_i^2\sum_{j<k}u(r_{jk})+
\left|\nablab_i\sum_{j<k}u(r_{jk})\right|^2\right]~,
\eq
where $V=V(\RR)$ is the total potential energy and
$r_{ij}=|\rr_{ij}|=|\rr_i-\rr_j|$. Then the 
antisymmetrized second order wave-function has the form in
Eq. (\ref{twfi}), which includes backflow (see the third term), which
is the correction inside the determinant and which affects the nodes,
and three-body boson-like correlations (see last term) which do not
affect the nodes.  

\begin{acknowledgments}
The MC simulations presented were carried out at the Center for High
Performance Computing (CHPC), CSIR Campus, 15 Lower Hope St.,
Rosebank, Cape Town, South Africa.
\end{acknowledgments}

\begin{thebibliography}{67}
\expandafter\ifx\csname natexlab\endcsname\relax\def\natexlab#1{#1}\fi
\expandafter\ifx\csname bibnamefont\endcsname\relax
  \def\bibnamefont#1{#1}\fi
\expandafter\ifx\csname bibfnamefont\endcsname\relax
  \def\bibfnamefont#1{#1}\fi
\expandafter\ifx\csname citenamefont\endcsname\relax
  \def\citenamefont#1{#1}\fi
\expandafter\ifx\csname url\endcsname\relax
  \def\url#1{\texttt{#1}}\fi
\expandafter\ifx\csname urlprefix\endcsname\relax\def\urlprefix{URL }\fi
\providecommand{\bibinfo}[2]{#2}
\providecommand{\eprint}[2][]{\url{#2}}

\bibitem[{\citenamefont{{N. W. Ashcroft and N. D.
  Mermin}}(1976)}]{Ashcroft-Mermin}
\bibinfo{author}{\bibnamefont{{N. W. Ashcroft and N. D. Mermin}}},
  \emph{\bibinfo{title}{Solid State Physics}} (\bibinfo{publisher}{Harcourt,
  Inc.}, \bibinfo{address}{Forth Worth}, \bibinfo{year}{1976}).

\bibitem[{\citenamefont{{S. L. Shapiro and S. A.
  Teukolsky}}(1983)}]{Shapiro-Teukolsky}
\bibinfo{author}{\bibnamefont{{S. L. Shapiro and S. A. Teukolsky}}},
  \emph{\bibinfo{title}{Black Holes, White Dwarfs, and Neutron Stars. The
  Physics of Compact Objects}} (\bibinfo{publisher}{John Wiley \& Sons, Inc.},
  \bibinfo{address}{Germany}, \bibinfo{year}{1983}).

\bibitem[{\citenamefont{{W. Zhu, J. Toulouse, A. Savin, and G.
  Angyan}}(2010)}]{Zhu2010}
\bibinfo{author}{\bibnamefont{{W. Zhu, J. Toulouse, A. Savin, and G. Angyan}}},
  \bibinfo{journal}{J. Chem. Phys.} \textbf{\bibinfo{volume}{132}},
  \bibinfo{pages}{244108} (\bibinfo{year}{2010}).

\bibitem[{\citenamefont{{J. Toulouse, P. Gori-Giorgi, and A.
  Savin}}(2005)}]{Toulouse2005}
\bibinfo{author}{\bibnamefont{{J. Toulouse, P. Gori-Giorgi, and A. Savin}}},
  \bibinfo{journal}{Theor. Chem. Acc.} \textbf{\bibinfo{volume}{114}},
  \bibinfo{pages}{305} (\bibinfo{year}{2005}).

\bibitem[{\citenamefont{{P. Gori-Giorgi and A. Savin}}(2006)}]{Gori2006}
\bibinfo{author}{\bibnamefont{{P. Gori-Giorgi and A. Savin}}},
  \bibinfo{journal}{Phys. Rev. A} \textbf{\bibinfo{volume}{3}},
  \bibinfo{pages}{032506} (\bibinfo{year}{2006}).

\bibitem[{\citenamefont{{D. M. Ceperley}}(1995)}]{Ceperley1995}
\bibinfo{author}{\bibnamefont{{D. M. Ceperley}}}, \bibinfo{journal}{Rev. Mod.
  Phys.} \textbf{\bibinfo{volume}{67}}, \bibinfo{pages}{279}
  (\bibinfo{year}{1995}).

\bibitem[{\citenamefont{{D. M. Ceperley}}(1996)}]{Ceperley1996}
\bibinfo{author}{\bibnamefont{{D. M. Ceperley}}}, in
  \emph{\bibinfo{booktitle}{Monte Carlo and Molecular Dynamics of Condensed
  Matter Systems}}, edited by \bibinfo{editor}{\bibnamefont{{K. Binder and G.
  Ciccotti}}} (\bibinfo{publisher}{Editrice Compositori},
  \bibinfo{address}{Bologna, Italy}, \bibinfo{year}{1996}).

\bibitem[{\citenamefont{{D. M. Ceperley and B. J. Alder}}(1980)}]{Ceperley1980}
\bibinfo{author}{\bibnamefont{{D. M. Ceperley and B. J. Alder}}},
  \bibinfo{journal}{Phys. Rev. Lett.} \textbf{\bibinfo{volume}{45}},
  \bibinfo{pages}{566} (\bibinfo{year}{1980}).

\bibitem[{\citenamefont{{L. Zecca, P. Gori-Giorgi, S. Moroni, and G. B.
  Bachelet}}(2004)}]{Zecca2004}
\bibinfo{author}{\bibnamefont{{L. Zecca, P. Gori-Giorgi, S. Moroni, and G. B.
  Bachelet}}}, \bibinfo{journal}{Phys. Rev. B} \textbf{\bibinfo{volume}{70}},
  \bibinfo{pages}{205127} (\bibinfo{year}{2004}).

\bibitem[{\citenamefont{{S. Paziani, S. Moroni, P. Gori-Giorgi, and G. B.
  Bachelet}}(2006)}]{Paziani2006}
\bibinfo{author}{\bibnamefont{{S. Paziani, S. Moroni, P. Gori-Giorgi, and G. B.
  Bachelet}}}, \bibinfo{journal}{Phys. Rev. B} \textbf{\bibinfo{volume}{73}},
  \bibinfo{pages}{155111} (\bibinfo{year}{2006}).

\bibitem[{\citenamefont{{R. Fantoni}}(2013)}]{Fantoni13}
\bibinfo{author}{\bibnamefont{{R. Fantoni}}}, \bibinfo{journal}{Solid State
  Communications} \textbf{\bibinfo{volume}{159}}, \bibinfo{pages}{106}
  (\bibinfo{year}{2013}).

\bibitem[{\citenamefont{{J. Kolorenc and L. Mitas}}(2011)}]{Kolorenc2011}
\bibinfo{author}{\bibnamefont{{J. Kolorenc and L. Mitas}}},
  \bibinfo{journal}{Rep. Prog. Phys.} \textbf{\bibinfo{volume}{74}},
  \bibinfo{pages}{026502} (\bibinfo{year}{2011}).

\bibitem[{\citenamefont{{W. M. C. Foulkes, L. Mitas, R. J. Needs, and G.
  Rajagopal}}(2001)}]{Foulkes2001}
\bibinfo{author}{\bibnamefont{{W. M. C. Foulkes, L. Mitas, R. J. Needs, and G.
  Rajagopal}}}, \bibinfo{journal}{Rev. Mod. Phys.}
  \textbf{\bibinfo{volume}{73}}, \bibinfo{pages}{33} (\bibinfo{year}{2001}).

\bibitem[{\citenamefont{{Y. Kwon, D. M. Ceperley, and R. M.
  Martin}}(1998)}]{Kwon1998}
\bibinfo{author}{\bibnamefont{{Y. Kwon, D. M. Ceperley, and R. M. Martin}}},
  \bibinfo{journal}{Phys. Rev. B} \textbf{\bibinfo{volume}{58}},
  \bibinfo{pages}{6800} (\bibinfo{year}{1998}).

\bibitem[{\citenamefont{{J. Toulouse, R. Assaraf, and C. J.
  Umrigar}}(2007)}]{Toulouse2007}
\bibinfo{author}{\bibnamefont{{J. Toulouse, R. Assaraf, and C. J. Umrigar}}},
  \bibinfo{journal}{J. Chem. Phys.} \textbf{\bibinfo{volume}{126}},
  \bibinfo{pages}{244112} (\bibinfo{year}{2007}).

\bibitem[{\citenamefont{{C. Lin, F. H. Zong, and D. M.
  Ceperley}}(2001)}]{Lin2001}
\bibinfo{author}{\bibnamefont{{C. Lin, F. H. Zong, and D. M. Ceperley}}},
  \bibinfo{journal}{Phys. Rev. E} \textbf{\bibinfo{volume}{64}},
  \bibinfo{pages}{016702} (\bibinfo{year}{2001}).

\bibitem[{\citenamefont{{S. Chiesa, D. M. Ceperley, R. M. Martin, and M.
  Holzmann}}(2006)}]{Chiesa2006}
\bibinfo{author}{\bibnamefont{{S. Chiesa, D. M. Ceperley, R. M. Martin, and M.
  Holzmann}}}, \bibinfo{journal}{Phys. Rev. Lett.}
  \textbf{\bibinfo{volume}{97}}, \bibinfo{pages}{076404}
  (\bibinfo{year}{2006}).

\bibitem[{\citenamefont{{W. L. McMillan}}(1965)}]{McMillan1965}
\bibinfo{author}{\bibnamefont{{W. L. McMillan}}}, \bibinfo{journal}{Phys. Rev.
  A} \textbf{\bibinfo{volume}{138}}, \bibinfo{pages}{442}
  (\bibinfo{year}{1965}).

\bibitem[{\citenamefont{{M. H. Kalos, D. Levesque, and L.
  Verlet}}(1974)}]{Kalos1974}
\bibinfo{author}{\bibnamefont{{M. H. Kalos, D. Levesque, and L. Verlet}}},
  \bibinfo{journal}{Phys. Rev. A} \textbf{\bibinfo{volume}{9}},
  \bibinfo{pages}{2178} (\bibinfo{year}{1974}).

\bibitem[{\citenamefont{Hockney and Eastwood}(1981)}]{Hockney}
\bibinfo{author}{\bibfnamefont{R.~W.} \bibnamefont{Hockney}} \bibnamefont{and}
  \bibinfo{author}{\bibfnamefont{J.~W.} \bibnamefont{Eastwood}},
  \emph{\bibinfo{title}{{"Computer Simulation Using Particles"}}}
  (\bibinfo{publisher}{McGraw-Hill}, \bibinfo{year}{1981}).

\bibitem[{\citenamefont{Allen and Tildesley}(1987)}]{Allen-Tildesley}
\bibinfo{author}{\bibfnamefont{M.~P.} \bibnamefont{Allen}} \bibnamefont{and}
  \bibinfo{author}{\bibfnamefont{D.~J.} \bibnamefont{Tildesley}},
  \emph{\bibinfo{title}{Computer Simulation of Liquids}}
  (\bibinfo{publisher}{Clarendon Press}, \bibinfo{address}{Oxford},
  \bibinfo{year}{1987}).

\bibitem[{\citenamefont{{D. Frenkel and B. Smit}}(1996)}]{Frenkel-Smit}
\bibinfo{author}{\bibnamefont{{D. Frenkel and B. Smit}}},
  \emph{\bibinfo{title}{Understanding Molecular Simulation}}
  (\bibinfo{publisher}{Academic Press}, \bibinfo{address}{San Diego},
  \bibinfo{year}{1996}).

\bibitem[{\citenamefont{Ceperley}(1991)}]{Ceperley1991}
\bibinfo{author}{\bibfnamefont{D.~M.} \bibnamefont{Ceperley}},
  \bibinfo{journal}{J. Stat. Phys.} \textbf{\bibinfo{volume}{63}},
  \bibinfo{pages}{1237} (\bibinfo{year}{1991}).

\bibitem[{\citenamefont{{D. M. Ceperley and M. H. Kalos}}(1979)}]{Ceperley1979}
\bibinfo{author}{\bibnamefont{{D. M. Ceperley and M. H. Kalos}}}, in
  \emph{\bibinfo{booktitle}{Monte Carlo Methods in Statistical Physics}},
  edited by \bibinfo{editor}{\bibfnamefont{K.}~\bibnamefont{Binder}}
  (\bibinfo{publisher}{Springer-Verlag}, \bibinfo{address}{Heidelberg},
  \bibinfo{year}{1979}), p. \bibinfo{pages}{145}.

\bibitem[{\citenamefont{{K. S. Liu, M. H. Kalos, and G. V.
  Chester}}(1974)}]{Kalos1974b}
\bibinfo{author}{\bibnamefont{{K. S. Liu, M. H. Kalos, and G. V. Chester}}},
  \bibinfo{journal}{Phys. Rev. A} \textbf{\bibinfo{volume}{10}},
  \bibinfo{pages}{303} (\bibinfo{year}{1974}).

\bibitem[{\citenamefont{{R. N. Barnett, P. J. Reynolds, and W. A. Lester,
  Jr.}}(1991)}]{Barnett1991}
\bibinfo{author}{\bibnamefont{{R. N. Barnett, P. J. Reynolds, and W. A. Lester,
  Jr.}}}, \bibinfo{journal}{J. Comp. Phys.} \textbf{\bibinfo{volume}{96}},
  \bibinfo{pages}{258} (\bibinfo{year}{1991}).

\bibitem[{\citenamefont{{S. Baroni and S. Moroni}}(1999)}]{Baroni1999}
\bibinfo{author}{\bibnamefont{{S. Baroni and S. Moroni}}},
  \bibinfo{journal}{Phys. Rev. Lett.} \textbf{\bibinfo{volume}{82}},
  \bibinfo{pages}{4745} (\bibinfo{year}{1999}).

\bibitem[{\citenamefont{{C. J. Umrigar, M. P. Nightingale, and K. J.
  Runge}}(1993)}]{Umrigar1993}
\bibinfo{author}{\bibnamefont{{C. J. Umrigar, M. P. Nightingale, and K. J.
  Runge}}}, \bibinfo{journal}{J. Chem. Phys.} \textbf{\bibinfo{volume}{99}},
  \bibinfo{pages}{2865} (\bibinfo{year}{1993}).

\bibitem[{\citenamefont{{R. Assaraf and M. Caffarel}}(2003)}]{Assaraf2003}
\bibinfo{author}{\bibnamefont{{R. Assaraf and M. Caffarel}}},
  \bibinfo{journal}{J. Chem. Phys.} \textbf{\bibinfo{volume}{119}},
  \bibinfo{pages}{10536} (\bibinfo{year}{2003}).

\bibitem[{\citenamefont{{R. Gaudoin and J. M. Pitarke}}(2007)}]{Gaudoin2007}
\bibinfo{author}{\bibnamefont{{R. Gaudoin and J. M. Pitarke}}},
  \bibinfo{journal}{Phys. Rev. Lett.} \textbf{\bibinfo{volume}{99}},
  \bibinfo{pages}{126406} (\bibinfo{year}{2007}).

\bibitem[{\citenamefont{Wigner}(1934)}]{Wigner1934}
\bibinfo{author}{\bibfnamefont{E.}~\bibnamefont{Wigner}},
  \bibinfo{journal}{Phys. Rev.} \textbf{\bibinfo{volume}{46}},
  \bibinfo{pages}{1002} (\bibinfo{year}{1934}).

\bibitem[{\citenamefont{{A. J. Leggett}}(1975)}]{Leggett1975}
\bibinfo{author}{\bibnamefont{{A. J. Leggett}}}, \bibinfo{journal}{Rev. Mod.
  Phys.} \textbf{\bibinfo{volume}{47}}, \bibinfo{pages}{331}
  (\bibinfo{year}{1975}).

\bibitem[{\citenamefont{{G. F. Giuliani and G.
  Vignale}}(2005)}]{Giuliani-Vignale}
\bibinfo{author}{\bibnamefont{{G. F. Giuliani and G. Vignale}}},
  \emph{\bibinfo{title}{Quantum Theory of the Electron Liquid}}
  (\bibinfo{publisher}{Cambridge University Press},
  \bibinfo{address}{Cambridge}, \bibinfo{year}{2005}).

\bibitem[{\citenamefont{{V. Natoli and D. M. Ceperley}}(1995)}]{Natoli1995}
\bibinfo{author}{\bibnamefont{{V. Natoli and D. M. Ceperley}}},
  \bibinfo{journal}{Comput. Phys.} \textbf{\bibinfo{volume}{117}},
  \bibinfo{pages}{171} (\bibinfo{year}{1995}).

\bibitem[{\citenamefont{{N. H. March and M. P. Tosi}}(1984)}]{March-Tosi}
\bibinfo{author}{\bibnamefont{{N. H. March and M. P. Tosi}}},
  \emph{\bibinfo{title}{Coulomb Liquids}} (\bibinfo{publisher}{Academic Press},
  \bibinfo{address}{London}, \bibinfo{year}{1984}).

\bibitem[{\citenamefont{Anderson}(1976)}]{Anderson1976}
\bibinfo{author}{\bibfnamefont{J.~B.} \bibnamefont{Anderson}},
  \bibinfo{journal}{J. Chem. Phys.} \textbf{\bibinfo{volume}{65}},
  \bibinfo{pages}{4121} (\bibinfo{year}{1976}).

\bibitem[{\citenamefont{{J. M. Hammersley and D. C.
  Handscomb}}(1964)}]{Hammersley}
\bibinfo{author}{\bibnamefont{{J. M. Hammersley and D. C. Handscomb}}},
  \emph{\bibinfo{title}{Monte Calro Methods}} (\bibinfo{publisher}{Chapman and
  Hall}, \bibinfo{address}{London}, \bibinfo{year}{1964}), \bibinfo{note}{pp.
  57-59}.

\bibitem[{\citenamefont{{M. H. Kalos and P. A.
  Whitlock}}(2008)}]{Kalos-Whitlock}
\bibinfo{author}{\bibnamefont{{M. H. Kalos and P. A. Whitlock}}},
  \emph{\bibinfo{title}{Monte Carlo Methods}} (\bibinfo{publisher}{Wiley-Vch
  Verlag GmbH \& Co.}, \bibinfo{address}{Germany}, \bibinfo{year}{2008}).

\bibitem[{\citenamefont{{L. D. Landau and E. M. Lifshitz}}(1977)}]{LandauQM}
\bibinfo{author}{\bibnamefont{{L. D. Landau and E. M. Lifshitz}}},
  \emph{\bibinfo{title}{{Quantum Mechanics. Non-relativistic Theory}}},
  vol.~\bibinfo{volume}{3} (\bibinfo{publisher}{Pergamon Press},
  \bibinfo{year}{1977}), \bibinfo{edition}{3rd} ed., \bibinfo{note}{course of
  Theoretical Physics. Eq. (11.16).}

\bibitem[{Bij()}]{Bijl1940}
\bibinfo{note}{A. Bijl, Physica {\bf 7}, 869 (1940); R. B. Dingle, Philos. Mag.
  {\bf 40}, 573 (1949); R. Jastrow, Phys. Rev. {\bf 98}, 1479 (1955)}.

\bibitem[{\citenamefont{{R. P. Feynman}}(1972)}]{Feynman}
\bibinfo{author}{\bibnamefont{{R. P. Feynman}}},
  \emph{\bibinfo{title}{Statistical Mechanics: A Set of Lectures}}
  (\bibinfo{publisher}{W. A. Benjamin Inc.}, \bibinfo{address}{London,
  Amsterdam, Don Mills, Sydney, Tokyo}, \bibinfo{year}{1972}),
  \bibinfo{note}{section 9.6}.

\bibitem[{\citenamefont{{T. Gaskell}}(1961)}]{Gaskell61}
\bibinfo{author}{\bibnamefont{{T. Gaskell}}}, \bibinfo{journal}{Proc. Phys.
  Soc.} \textbf{\bibinfo{volume}{77}}, \bibinfo{pages}{1182}
  (\bibinfo{year}{1961}).

\bibitem[{\citenamefont{{T. Gaskell}}(1962)}]{Gaskell62}
\bibinfo{author}{\bibnamefont{{T. Gaskell}}}, \bibinfo{journal}{Proc. Phys.
  Soc.} \textbf{\bibinfo{volume}{80}}, \bibinfo{pages}{1091}
  (\bibinfo{year}{1962}).

\bibitem[{\citenamefont{{D. M. Ceperley}}(2004)}]{Ceperley2004}
\bibinfo{author}{\bibnamefont{{D. M. Ceperley}}}, in
  \emph{\bibinfo{booktitle}{Proceedings of the International School of Physics
  Enrico Fermi}}, edited by \bibinfo{editor}{\bibnamefont{{G. F. Giuliani and
  G. Vignale}}} (\bibinfo{publisher}{IOS Press}, \bibinfo{address}{Amsterdam},
  \bibinfo{year}{2004}), pp. \bibinfo{pages}{3--42}, \bibinfo{note}{course
  CLVII}.

\bibitem[{rpa()}]{rpanote}
\bibinfo{note}{Note that the probability distribution in a variational
  calculation is (from Eq. (\ref{twf})) $\Psi^2(\RR)\propto
  D^2(\RR)\exp[-2U(\RR)]$ with $U(\RR)=\sum_{i<j}u(r_{ij})$. Then if one
  formally writes $D^2(\RR)=\exp[-2W(\RR)]$, $\Psi^2$ becomes the probability
  distribution for a classical fluid with potential $W+U$ at an inverse
  temperature $\beta=2$. Then one sees that with the choice ${\cal B}=2$, Eq
  (\ref{crpa}) coincides with the well known Random Phase Approximation in the
  theory of classical fluids (see Ref. \onlinecite{Hansen} Section 6.5) where
  $W$ is the potential of the reference fluid and $U$ the perturbation.}

\bibitem[{\citenamefont{{D. Ceperley}}(1978)}]{Ceperley78}
\bibinfo{author}{\bibnamefont{{D. Ceperley}}}, \bibinfo{journal}{Phys. Rev. B}
  \textbf{\bibinfo{volume}{18}}, \bibinfo{pages}{3126} (\bibinfo{year}{1978}).

\bibitem[{\citenamefont{{Y. Kwon, D. M. Ceperley, and R. M.
  Martin}}(1993)}]{Kwon1993}
\bibinfo{author}{\bibnamefont{{Y. Kwon, D. M. Ceperley, and R. M. Martin}}},
  \bibinfo{journal}{Phys. Rev. B} \textbf{\bibinfo{volume}{48}},
  \bibinfo{pages}{12037} (\bibinfo{year}{1993}).

\bibitem[{\citenamefont{{R. P. Feynman and M. Cohen}}(1956)}]{Feynman1956}
\bibinfo{author}{\bibnamefont{{R. P. Feynman and M. Cohen}}},
  \bibinfo{journal}{Phys. Rev.} \textbf{\bibinfo{volume}{102}},
  \bibinfo{pages}{1189} (\bibinfo{year}{1956}).

\bibitem[{\citenamefont{{R. M. Panoff and J. Carlson}}(1989)}]{Panoff1989}
\bibinfo{author}{\bibnamefont{{R. M. Panoff and J. Carlson}}},
  \bibinfo{journal}{Phys. Rev. Lett.} \textbf{\bibinfo{volume}{62}},
  \bibinfo{pages}{1130} (\bibinfo{year}{1989}).

\bibitem[{\citenamefont{{T. L. Hill}}(1956)}]{Hill}
\bibinfo{author}{\bibnamefont{{T. L. Hill}}}, \emph{\bibinfo{title}{Statistical
  Mechanics}} (\bibinfo{publisher}{McGraw-Hill}, \bibinfo{address}{New York},
  \bibinfo{year}{1956}).

\bibitem[{\citenamefont{{E. Feenberg}}(1967)}]{Feenberg}
\bibinfo{author}{\bibnamefont{{E. Feenberg}}}, \emph{\bibinfo{title}{Theory of
  Quantum Fluids}} (\bibinfo{publisher}{Academic Press}, \bibinfo{year}{1967}).

\bibitem[{\citenamefont{{Ph. A. Martin}}(1988)}]{Martin88}
\bibinfo{author}{\bibnamefont{{Ph. A. Martin}}}, \bibinfo{journal}{Rev. Mod.
  Phys.} \textbf{\bibinfo{volume}{60}}, \bibinfo{pages}{1075}
  (\bibinfo{year}{1988}).

\bibitem[{\citenamefont{Kimball}(1973)}]{Kimball1973}
\bibinfo{author}{\bibfnamefont{J.~C.} \bibnamefont{Kimball}},
  \bibinfo{journal}{Phys. Rev. A} \textbf{\bibinfo{volume}{7}},
  \bibinfo{pages}{1648} (\bibinfo{year}{1973}).

\bibitem[{\citenamefont{{A. K. Rajagopal, J. C. Kimball, and M.
  Banerjee}}(1978)}]{Rajagopal1978}
\bibinfo{author}{\bibnamefont{{A. K. Rajagopal, J. C. Kimball, and M.
  Banerjee}}}, \bibinfo{journal}{Phys. Rev. B} \textbf{\bibinfo{volume}{18}},
  \bibinfo{pages}{2339} (\bibinfo{year}{1978}).

\bibitem[{\citenamefont{{M. Hoffmann-Ostenhof, T. Hofmann-Ostenhof, and H.
  Stremnitzer}}(1992)}]{Hoffmann1992}
\bibinfo{author}{\bibnamefont{{M. Hoffmann-Ostenhof, T. Hofmann-Ostenhof, and
  H. Stremnitzer}}}, \bibinfo{journal}{Phys. Rev. Lett.}
  \textbf{\bibinfo{volume}{68}}, \bibinfo{pages}{3857} (\bibinfo{year}{1992}).

\bibitem[{\citenamefont{{J. P. Hansen and I. R. McDonald}}(1986)}]{Hansen}
\bibinfo{author}{\bibnamefont{{J. P. Hansen and I. R. McDonald}}},
  \emph{\bibinfo{title}{Theory of simple liquids}}
  (\bibinfo{publisher}{Academic Press}, \bibinfo{address}{London},
  \bibinfo{year}{1986}), \bibinfo{edition}{2nd} ed.

\bibitem[{sta()}]{static}
\bibinfo{note}{Note that, unlike in the classical case, in quantum statistical
  physics even the linear response to a {\sl static} perturbation requires the
  use of imaginary time correlation functions \cite{Martin88}.}

\bibitem[{\citenamefont{{L. van Hove}}(1954)}]{vanHove54}
\bibinfo{author}{\bibnamefont{{L. van Hove}}}, \bibinfo{journal}{Phys. Rev.}
  \textbf{\bibinfo{volume}{95}}, \bibinfo{pages}{249} (\bibinfo{year}{1954}).

\bibitem[{\citenamefont{{D. Pines and P. Nozi\`eres}}(1966)}]{Pines}
\bibinfo{author}{\bibnamefont{{D. Pines and P. Nozi\`eres}}},
  \emph{\bibinfo{title}{Theory of Quantum Liquids}}
  (\bibinfo{publisher}{Benjamin}, \bibinfo{address}{New York},
  \bibinfo{year}{1966}).

\bibitem[{\citenamefont{{J. Lindhard}}(1954)}]{Lindhard54}
\bibinfo{author}{\bibnamefont{{J. Lindhard}}}, \bibinfo{journal}{Mat.-Fys.
  Medd.} \textbf{\bibinfo{volume}{28}} (\bibinfo{year}{1954}).

\bibitem[{\citenamefont{{M. P. Tosi}}(1999)}]{Tosi1999}
\bibinfo{author}{\bibnamefont{{M. P. Tosi}}}, in
  \emph{\bibinfo{booktitle}{Electron Correlation in the Solid State}}, edited
  by \bibinfo{editor}{\bibnamefont{{N. H. March}}}
  (\bibinfo{publisher}{Imperial College Press}, \bibinfo{address}{London},
  \bibinfo{year}{1999}), chap.~\bibinfo{chapter}{1}, pp.
  \bibinfo{pages}{1--42}.

\bibitem[{\citenamefont{{A. Alastuey and Ph. A. Martin}}(1985)}]{Alastuey1985}
\bibinfo{author}{\bibnamefont{{A. Alastuey and Ph. A. Martin}}},
  \bibinfo{journal}{J. Stat. Phys.} \textbf{\bibinfo{volume}{39}},
  \bibinfo{pages}{405} (\bibinfo{year}{1985}).

\bibitem[{\citenamefont{Lighthill}(1959)}]{Lighthill}
\bibinfo{author}{\bibfnamefont{M.~J.} \bibnamefont{Lighthill}},
  \emph{\bibinfo{title}{Introduction to Fourier Analysis and Generalized
  Functions}} (\bibinfo{publisher}{Cambridge University Press},
  \bibinfo{year}{1959}), \bibinfo{note}{theorem 19}.

\bibitem[{shi()}]{shift}
\bibinfo{note}{Note that with the given choice of $Q$ we obtain $\langle\Delta
  I^{ZV}_{\sigma,\sigma^\prime}(r,\RR)\rangle_{\Psi^2}$
  $=-\int_{\partial\Omega^N}\Psi^2(\RR)\nablab
  Q_{\sigma,\sigma^\prime}(r,\RR)\cdot d\SSS/r_s^2$ $=-\Omega n_\sigma
  n_{\sigma^\prime}$, for all $r$ with $\rr\in\Omega$, instead of zero as
  normally expected. This is ultimately related to the behavior of the
  auxiliary function $\Psi^\prime=Q\Psi$ on the border of $\Omega^N$.}

\bibitem[{\citenamefont{{G. Ortiz and P. Ballone}}(1994)}]{Ortiz1994}
\bibinfo{author}{\bibnamefont{{G. Ortiz and P. Ballone}}},
  \bibinfo{journal}{Phys. Rev. B} \textbf{\bibinfo{volume}{50}},
  \bibinfo{pages}{1391} (\bibinfo{year}{1994}).

\bibitem[{\citenamefont{{M. Kac}}(1959)}]{Kac59}
\bibinfo{author}{\bibnamefont{{M. Kac}}}, \emph{\bibinfo{title}{Probability and
  Related Topics in Physical Sciences}} (\bibinfo{publisher}{Interscience
  Publisher Inc.}, \bibinfo{address}{New York}, \bibinfo{year}{1959}).

\bibitem[{Kac(1951)}]{Kac51}
\emph{\bibinfo{title}{Proceedings of the Second Berkeley Symposium on
  Probability and Statistics}} (\bibinfo{publisher}{University of California
  Press}, \bibinfo{address}{Berkeley}, \bibinfo{year}{1951}),
  \bibinfo{note}{sec. 3}.

\end{thebibliography}

\end{document}